\documentclass[aps,prd,superscriptaddress,nofootinbib,floatfix]{revtex4-1}

\usepackage{graphicx}
\usepackage{amsmath}
\usepackage{amssymb}
\usepackage{yhmath}
\usepackage{extarrows}
\usepackage[colorlinks,citecolor=blue,linkcolor=blue,urlcolor=blue,anchorcolor=blue]{hyperref}
\allowdisplaybreaks
\makeatletter

\@addtoreset{equation}{section}
\makeatother

\newcommand\etal{{\it et~al.}}

\begin{document}

\title{Image of the Schwarzschild black hole pierced
 by a cosmic string with a thin accretion disk}
\author{Changqing Liu}
\email[]{lcqliu2562@163.com,Corresponding author}
\author{Li Tang}
\email[]{2026002847@qq.com}
\affiliation{Department of Physics,Hunan University of Humanities Science and Technology, Loudi, Hunan 417000, P. R. China }
\author{Jiliang  Jing}
\email[]{jljing@hunnu.edu.cn}
\affiliation{Department of Physics, and Key Laboratory of Low Dimensional,Quantum Structures and Quantum Control of Ministry of Education, Hunan Normal University,  Changsha, Hunan 410081, P. R. China}
\date{\today}

\begin{abstract}
We study the optical appearance of a thin accretion disk around a Schwarzschild black hole pierced
 by a cosmic string with a semi-analytic method of Luminet\cite{Luminet1979}. Direct and secondary images with different parameters observed
 by a distant observer is plotted. The cosmic string parameter $s$ can modify the shape and size of the thin disk image. We calculate and plot the distribution of both redshift and observed flux as seen by distant observers at different inclination angles. Those distributions are dependent on the inclination angel of the observer and cosmic parameter $s$.
\end{abstract}
\maketitle

\section{Introduction}

In 2019, the Event Horizon Telescope (EHT) collaboration reported the images of the supermassive object at the center of the M87 galaxy\cite{EHT1}. Very recently, EHT collaboration have also released the first polarized images of the black hole M87* \cite{Narayan}. This works provide us lots of enlightening answers to probe general relativity, understanding the nature of accretion disk around the supermassive compact objects in the strong-gravity near-horizon regime and also test the other modified theories.

 Images of accretion disk around black holes have been one of the most fascinating topics of research in observational astronomy since the 1970s. A standard model of geometrically thin and optically thick accretion disk was proposed in Ref\cite{ss} and developed by
Novikov and Thorne\cite{Novikov}. Cunningham and Bardeen \cite{Cunningham:1972,Cunningham:1975} firstly calculated the optical appearance
of a star orbiting a Kerr black hole. The technique for the calculating photon geodesics and the images of thin accretion disk mainly include two methods: the semi-analytic method and
ray-tracing combining with radiative transfer method. Especially, Luminet\cite{Luminet1979} used the semi-analytic method to obtain the direct and secondary images of an accretion disk around Schwarzschild black hole. A unique feature of this method is that observed physics quantities such as radiation flux, redshift etc, are expressed in terms of impact parameter and elliptic integrals. Bao $\etal$ obtained
the optical appearance of a source orbiting on a relativistic eccentric orbit. Unlike the circular orbit, the images of the orbiting source on the eccentric orbit  is not closed due the
relativistic advance of the periastron\cite{BAO}.
On the other hand, a series of numerical ray-tracing codes(i.e \cite{Viergutz,speith,fanton,Broderic,dexter,Vincent,ynogkm,TMuller,PCunha} ) was adopted to simulating the accretion structures around black holes. The optical appearance and physical properties of thin accretion disk in a variety of background space-times  have also been investigated extensively in \cite{Takahashi,Gyul2019,Shaikh,bambi_2019d,Johannsen,gate, ali}. The polarized image of a synchrotron-emitting fluid ring orbiting
a Schwarzchild black hole with an analytical model is present in \cite{Narayan}. This model has been extended to Kerr black \cite{gelles} and the case of  the photons coupled to Weyl tensor \cite{zelin}. Polarization pattern of the image can provides a wealth of information about the electromagnetic emissions from the vicinity of black hole.

Cosmic strings originate from phase transition phase transitions in the early universe in
certain particle physics models. Different black holes pierced by cosmic strings are inevitably
formed in the early universe under a particular condition. Gravitational phenomenon in the background of black holes containing cosmic strings has gained a lot of
renewed interest over the past years. The analytical solutions of the equations of motion in the Schwarzschild and Kerr
spacetime pierced by black string were presented \cite{Hackmann,Hackmann1}. The deflection angle of the equatorial and quasi-equatorial lensed by the holes \cite{shaowen} are found to depend closely on the cosmic string parameter. The result shows that the cosmic string parameter has
a significant influence on the corresponding gravitational phenomenon.

The motivation of the present paper is to study a thin accretion disk surrounding the Schwarzschild black hole
pierced by a cosmic string\cite{aryal} in the Novikov-Thorne model\cite{Novikov}. Following the method in \cite{Luminet1979}, we will plot direct and secondary image observed by a distant observer. We also will depict the distribution of redshift and  flux of the images on
the photographic plate to analyse the possible deviation from the case of Schwarzschild black holes.

The paper is organized as follows. In section \ref{sec1}, we focus on the analysis of the photon geodesics in the equatorial plane of the Schwarzschild black hole pierced
 by a cosmic string. In section \ref{sec2}, we calculate the apparent shapes of circular rings orbiting a spherical black hole and
 the direct and secondary images are plotted in Figs.1-2. The images are compared to the Schwarzschild black hole. Finally,  we study the distribution of radiation flux and redshift of the thin accretion disk as seen by distant observers at various inclination angle angles. In section \ref{sec3}, we summarize our results. Through the paper we use
geometrical units with $G = c = 1$.

\section{ Null geodesics of the Schwarzschild black hole pierced
 by a cosmic string}\label{sec1}

The Schwarzschild black hole
pierced by a cosmic string, whose metric is
described as \cite{ aryal}
\begin{eqnarray}
ds^2=-(1-\frac{2m}{r})dt^2+(1-\frac{2m}{r})^{-1}dr^2
+r^2d\theta^2+s^2r^2\sin^2{\theta}d\varphi^2,\label{metric}
\end{eqnarray}
where $s$ is related to the linear mass density $\rho_s$ of the string
by $s=1-4\rho_s$ and is justified in the range $0<s\leq 1$ as
$\rho_s\ll 1$. The physical mass  $M$ of
the black hole is given as  $M=s m$\cite{Galtsov}.
Due to the spherical symmetry of the Schwarzschild black hole
pierced by a cosmic string we can restrict geodesics
to the equatorial plane($\theta=\frac{\pi}{2}$). The Lagrangian equation of the photon motion is given as
\begin{equation}
L = \frac{1}{2}g_{\mu \nu} \frac{d x^\mu} {d \lambda } \frac{d x^\nu}{d \lambda}=0.
\end{equation}
The energy $\tilde{E}$ and angular momentum $\tilde{L}$ of null test particles
\begin{eqnarray}
\tilde{E}=-g_{tt} \dot t ,\\
 \tilde{L}=g_{\varphi \varphi} \dot \varphi. \\ \nonumber
\end{eqnarray}
From the Euler-Lagrange equations for null geodesics, we get
\begin{equation}\label{eqv}
\frac{1}{s^4r^4}(\frac{dr}{d\varphi})^2+V(r)=\frac{1}{b^2},
\end{equation}
where $b=\tilde{L}/\tilde{E}$ is the impact parameter at infinity, and $V=\frac{1}{s^2r^2}(1-\frac{2M}{s r})$  denotes the effective potential of null geodesics. As $dV(r)/r=0$, the function $V(r)$ has its maximum at $r_c=3M/s$, its value is $V_c=1/(27M^2)$. Thus, there exists a critical impact parameter $b_c=3\sqrt{3}M$ such that, for $b>3\sqrt{3}M$, rays of light are deflected but not captured by the black hole, and for  $b<3\sqrt{3}M$, rays are captured.

Letting $u=1/r$, we rewrite the  Eq. (\ref{eqv}) as
\begin{equation}\label{eqvu}
(\frac{du}{d\varphi})^2=s(2Mu^3-su^2+\frac{s^3}{b^2})\equiv 2MsG(u)=2Ms(u-u_1)(u-u_2)(u-u_3).
\end{equation}
We are only interested in rays that can reach a distant observer's plane when $b>b_c$. It's not hard to see that the cubic polynomial $G(u)$ has only only real negative root $u_1$ and two distinct positive roots $u_2$ and $u_3$. The three roots, given by a periastron distance $P$ with the order $u_1<u_2<u_3$, are expressed as \cite{chandra}:
\begin{eqnarray}
u_1 = \frac{R -2M-Q}{4MP}, \quad  u_2
= \frac{1}{P}, \quad u_3 = \frac{R -2M+Q}{4M P}.
\end{eqnarray}
By comparing the coefficients in $G(u)$ to those in the original
polynomial in   Eq. (\ref{eqvu}), we obtain the following the quantities
\begin{eqnarray}
R&=&Ps,\\
Q^2&=&(R - 2 M) (R+ 6M),\\
b&=&\frac{s^{\frac{3}{2}}P^{\frac{3}{2}}}{\sqrt{P s-2M}}.
\end{eqnarray}
So, given a value of the periastron $P$, we can obtain the value for the
parameter $b$ at infinity , and conversely.

The bending angle of the light ray is given by (\cite{iyer}):
\begin{eqnarray}
\mu &=& 2 \int_0^{u_2} \frac{du}{\sqrt{2M s(u-u_1)(u-u_2)(u-u_3)}}
-\pi\\ && \nonumber \\
&=& \sqrt{\frac{2}{M s}}\int_0^{u_2} \frac{du}{\sqrt{(u-u_1)(u_2-u)(u_3-u)}}-\pi.
\end{eqnarray}

Split the above integral into two parts to make the lower limit equal to
the smallest root $u_1$:
\begin{eqnarray}
\label{eqmu}
\mu =\sqrt{\frac{2}{M s}} \left[ \int_{u_1}^{u_2}
\frac{du}{\sqrt{(u-u_1)(u_2-u)(u_3-u)}} \
- \ \int_{u_1}^0 \frac{du}{\sqrt{(u-u_1)(u_2-u)(u_3-u)}} \right] \ - \ \pi.
\end{eqnarray}
Now, the integrals in Eq. (\ref{eqmu}) can be realized as elliptic
integrals: \cite{Luminet1979, iyer,chandra, darwin}
\begin{eqnarray}
\mu =\sqrt{\frac{2}{M s}} \left[ \,
\frac{2\, F (\Psi_1, k)}{\sqrt{u_3 - u_1}}  \
- \ \frac{2\, F (\Psi_2, k)}{\sqrt{u_3 - u_1}} \right] \ - \ \pi,
\end{eqnarray}
where $F(\Psi_i,k)$ is an incomplete elliptic integral of the first kind with amplitudes
$$ \Psi_1 = \frac{\pi}{2},\quad \Psi_2 = \sin^{-1} \sqrt{\frac{-u_1}{u_2 - u_1}}, $$
with  $$ k^2= \frac{u_2 - u_1}{u_3 - u_1}. $$,
$$ \Psi_2 = \sin^{-1} \sqrt{\frac{Q  + 2 M -R}{Q
+ 6 M -R}}.  $$
Hence, the bending angle of the light ray  simplifies to
\begin{eqnarray}
\label{eq-exact-bangle}
\mu=4\sqrt{\frac{P}{Q s}} \left[K(k) \
- \ F(\Psi_2,k)\right]\ -\ \pi,
\end{eqnarray}
where $K(k)$ is the complete
elliptic integrals of the first kind.

\section{Optical appearance of the Schwarzschild black hole
pierced by a cosmic string with a thin accretion disk}\label{sec2}

In this section, We use Luminet's method\cite{Luminet1979} to study the optical appearance of the Schwarzschild black hole
pierced by a cosmic string with the Novikov-Thorne thin disk model \cite{Novikov,Page:1974}.

\subsection{The image and  coordinate system of the thin accretion disk}
\begin{figure}[h!]
    		\setlength{\tabcolsep}{ 0 pt }{\footnotesize\tt
		\begin{tabular}{ cc }
           \includegraphics[width=1\textwidth]{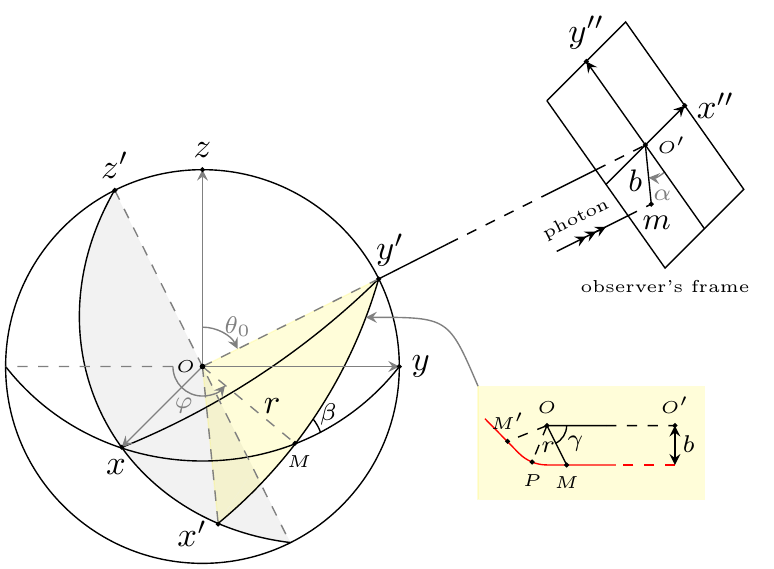}\\
           		\end{tabular}}
 \caption{\label{coord}\small The coordinate system of the disk \cite{Luminet1979},\cite{zhu}. }
\end{figure}

The coordinate system of the thin accretion disk is plotted in Fig. \ref{coord} and has been discussed in detail in the literature \cite{zhu,Luminet1979}.
The black hole and the disk is placed at point $o$ and the $\overline{xoy}$ plane, respectively. The observer lies in the fixed direction $\theta_0, \varphi=0$. Point $M$ means a disk element. In the bottom right subgraph, the red line is the photon trajectory, point $P$ is the perihelion of the photon trajectory with the impact parameter $b$.

We assume that the observer is at infinity and at rest in gravitational field of the Schwarzschild black hole
pierced by a cosmic string, then the polar distance from $m$ to $O'$ is exactly the impact parameter of the trajectory, and the polar angle $\alpha$ with "vertical" direction $O'Y''$ is the complement of dihedral angle between planes $OXY'$ and $OX'Y'$.
Rays emit from point $M$ with the coordinates $(r,\varphi)$ and reach point $m$ in the photographic plate $O'X''Y''$ with the coordinates $(b,\alpha)$. Thus, for a given emitter $M$, varying $\varphi$ from $0$ to $\pi$, the observer will obtain the two images of accretion disk around the black hole, a direct image
with the polar coordinates $(b^{(d)},\alpha)$ and secondary image with the coordinates ($b^{(s)},\alpha +\pi$)\cite{Luminet1979}. We can obtain the relationship among the deflection angle  from $M$ to the observer $\gamma$, the observer's inclination angle $\theta_0$ and $\alpha$ with the help of the sine theorem of spherical triangle \cite{zhu,Luminet1979}

\begin{equation}\label{eqrs}
  \cos\gamma=\frac{\cos\alpha}{\sqrt{(\cos^2\alpha+\cot^2\theta_0)}}.
\end{equation}
For the direct image of the accretion disk, formula (\ref{eqvu}) can be integrated explicitly
in terms of elliptical integrals \cite{Luminet1979,chandra,Muller:2009}.
\begin{eqnarray}\label{qrs}
  \gamma&=&\int_0^{1/r}\frac{d u}{\sqrt{G(u)}}\nonumber\\&=&2\sqrt{\frac{P}{Q s}}(F(\zeta_r,k)-F(\zeta_\infty,k)),
\end{eqnarray}
with
\begin{eqnarray}\label{qrw}
  k^2&=&\frac{Q-R+6M}{2Q},\nonumber\\
  sin^2\zeta_\infty&=&\frac{Q-R+2M}{Q-R+6M},\nonumber\\
  sin^2\zeta_r&=&\frac{Q-R+2M+4MP/r}{Q-R+6M},
\end{eqnarray}
where $F(\zeta_r,k)$ and $F(\zeta_\infty,k)$ are the elliptical integrals.

From Eq. (\ref{qrs}), we can express $r=r(\alpha,P)$ as function of $\alpha$ and periastron distance $P$
\begin{equation}\label{ew}
  \frac{1}{r}=-\frac{Q-R+2M}{4MP}+-\frac{Q-R+6M}{4MP}sn^2\left(\frac{\gamma}{2}\sqrt{\frac{Q s}{P}}+F(\zeta_\infty,k),k\right),
\end{equation}
where $sn\left(\frac{\gamma}{2}\sqrt{\frac{Q s}{P}}+F(\zeta_\infty,k),k\right)$ is  the Jacobi elliptic function.
 We can thus draw the iso-radial  curves ( corresponding to trajectories emitted at constant coordinate
 r from the black hole)  $b^{(d)}=b^{(d)}(\alpha)$ in polar coordinates, for a given angle $\theta_0$.

 Similarly, for the $(1+n)th$ order image of the accretion disk(a detailed discussion see \cite{Luminet1979}), formula (\ref{qrs}) take the
 following  the expressions
 \begin{eqnarray}\label{qrs}
  2n\pi-\gamma&=&2\sqrt{\frac{P}{Q s}}\left(2K(k)-(F(\zeta_r,k)-F(\zeta_\infty,k)\right),
\end{eqnarray}
here $K(k)$ is the complete elliptic integral.

\begin{figure}[h!]
    		\setlength{\tabcolsep}{ 0 pt }{\footnotesize\tt
		\begin{tabular}{ cc }
           \includegraphics[width=0.5\textwidth]{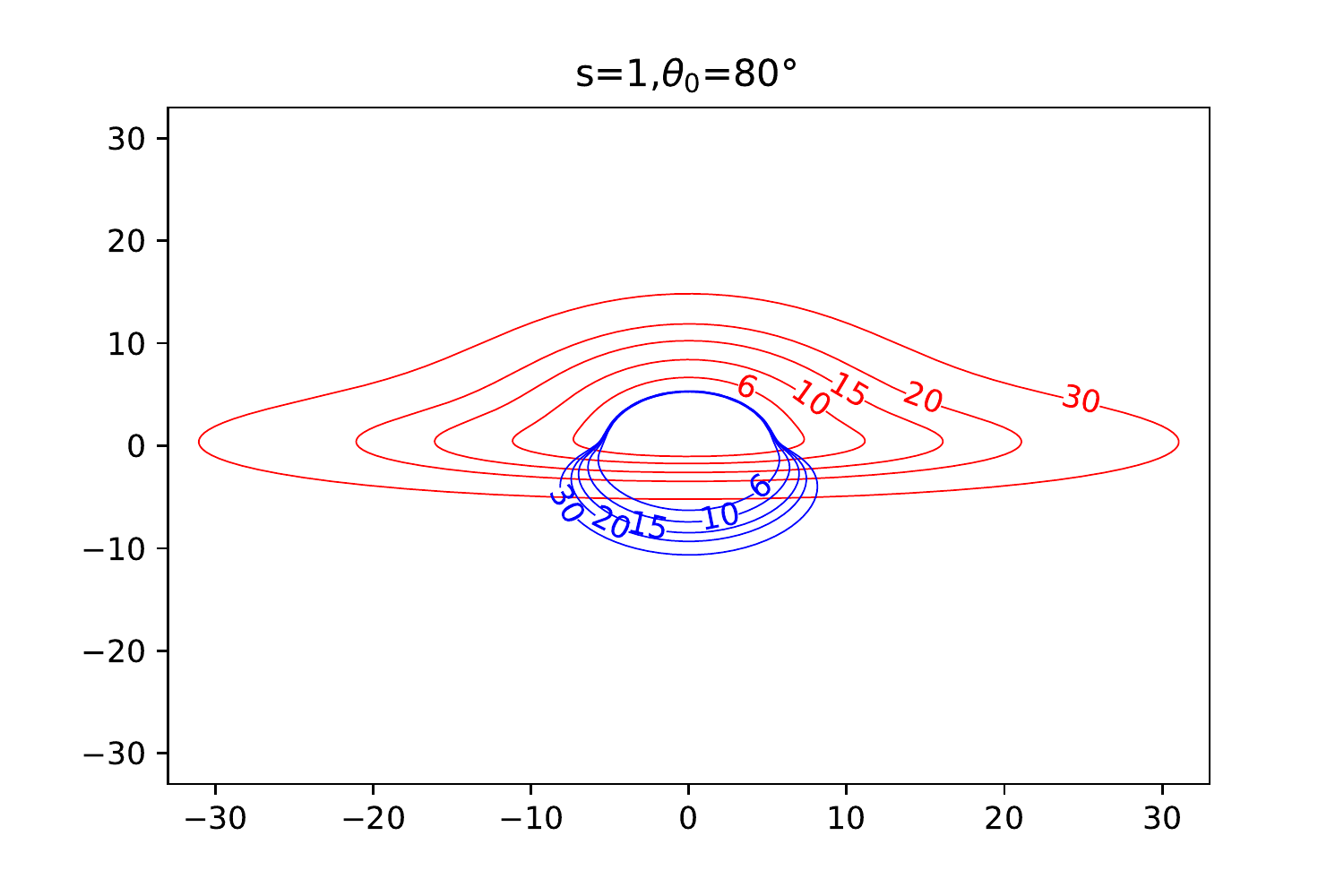}
           \includegraphics[width=0.5\textwidth]{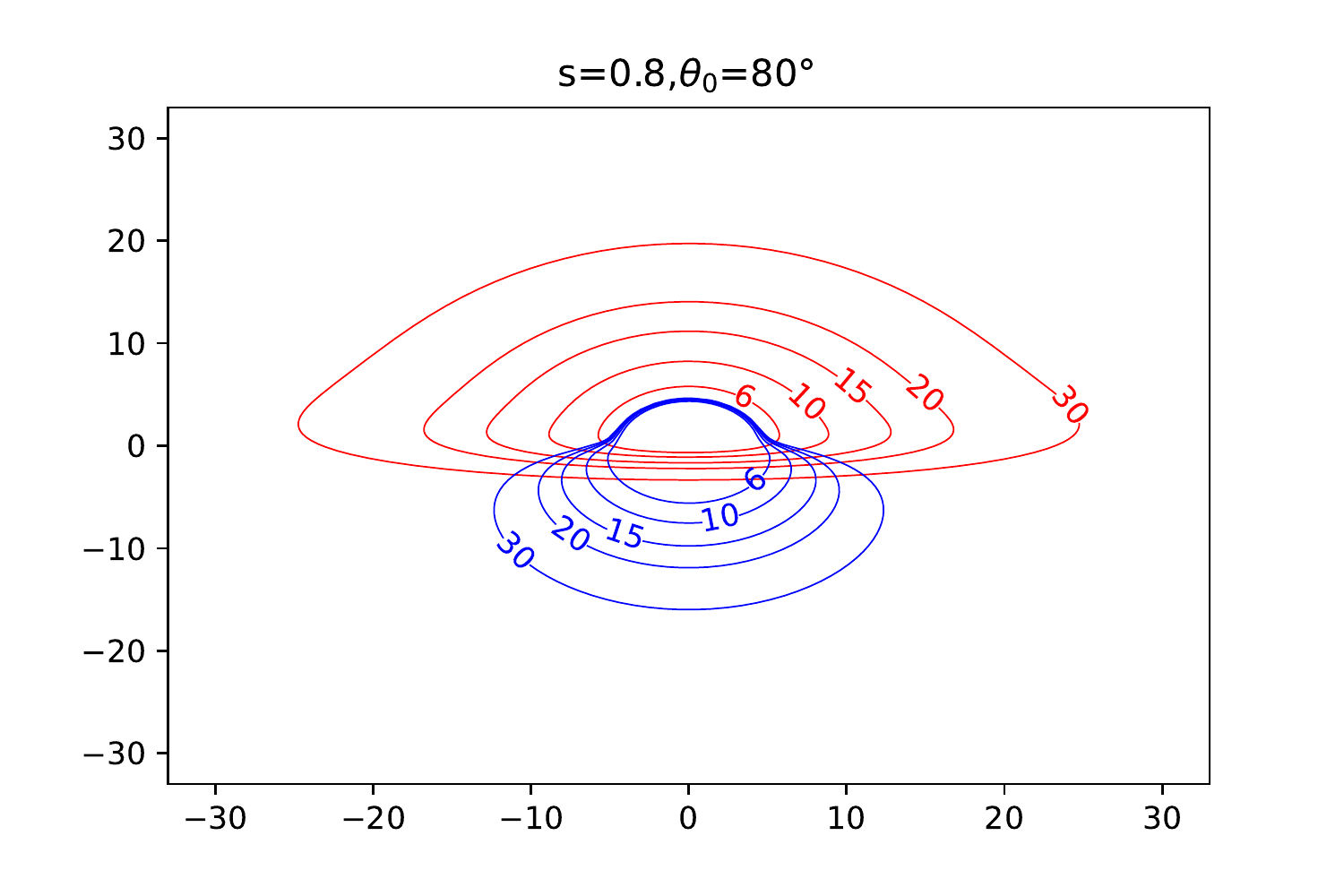} \\
            \includegraphics[width=0.5\textwidth]{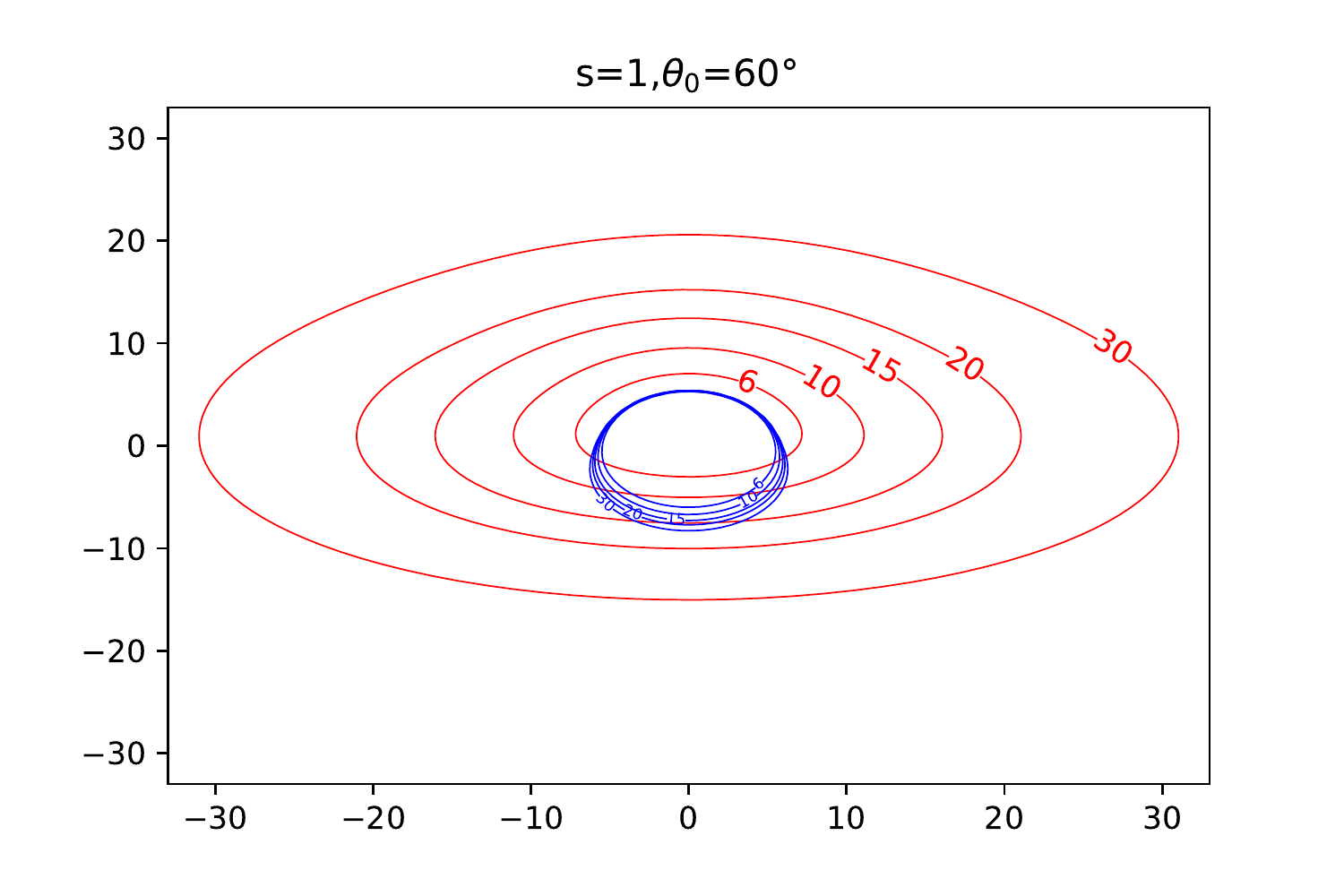}
             \includegraphics[width=0.5\textwidth]{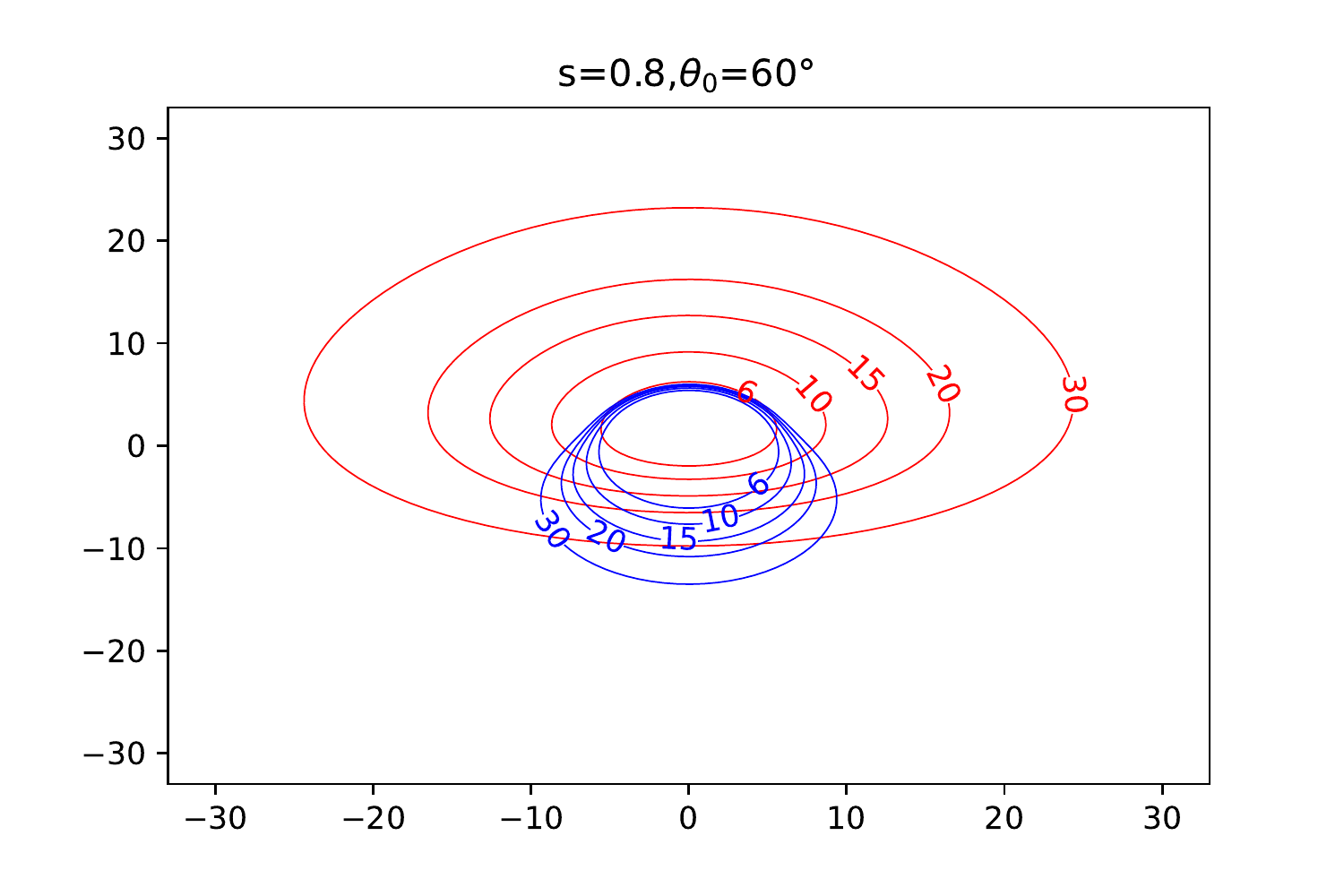} \\

           		\end{tabular}}
 \caption{\label{images}\small Direct and secondary image of the thin accretion disk around Schwazrschild black hole pierced by  a cosmic string with the inclination angles of the observer $\theta_0=60^\circ,80^\circ$. Red line is direct image of the disk , and blue line is secondary image of the disk. We set $M=1$. }
\end{figure}

Fig. ($\ref{images}$) present direct and secondary of circular rings orbiting around the Schwazrschild black hole pierced by a cosmic string at $r=6M,r=10M, r=15M, r=20M, r=30M$, for the observer's inclination angles $\theta=80^{\circ},60^{\circ}$ and the cosmic string parameter $s=1,0.8$, respectively. The direct images depicted in red line are generated by photons emitted in direction above the equatorial plane, while secondary images depicted in blue line correspond to photons heading in direction below the equatorial plane. We see that the direct images
 shrink( increase) with the decrease of the cosmic string parameter $s$ in the horizontal(vertical) direction, and the shape and size of the secondary images become much larger as the cosmic string parameter $s$ take the less value when the observers in high inclination angle $\theta_0$.

\subsection{Radiation from a thin accretion disk around the Schwarzschild black hole pierced
 by a cosmic string}

Until now we have considered the geometrical optics of a thin disk consisting of particles moving on circular equatorial orbits around the Schwarzschild black hole pierced
 by a cosmic string. In this section we will apply the Novikov-Thorne thin accretion disk model to analysis the
intrinsic radiation flux and red-shift from the disk surface. The thin accretion disk is non-self-gravitating; that is, the impact of
the disk's mass on the background metric is ignored, and the space-time is stationary, axisymmetric, asymptotically flat,
and reflection-symmetric with respect to the equatorial plane. The flux of the radiant energy over the disk can be expressed in terms of the
specific energy, angular momentum and of the angular
velocity of the black hole \cite{Novikov,Page:1974}

\begin{equation}\label{Fb}
\mathcal{F}_s(r)=-\frac{\dot{M}}{4\pi \sqrt{-g}}\frac{\Omega
_{,r}}{(E-\Omega
L)^{2}}\int_{r_{{in}}}^{r}(E-\Omega
L)L_{,r}dr,
\end{equation}
here, $g$ is the determinant of the induced metric in the equatorial plane, $r_{in}$ is the inner edge of the disk, $\dot M$ is the mass accretion rate, and $E$, $L$, and $\Omega$ are the energy, the angular momentum, and the angular velocity of the particles moving on a particular circular orbit. As a general static spherically symmetric metric  has the form
\begin{equation}\label{metric_st}
ds^2=g_{tt}\,dt^2+g_{\varphi\varphi}\,d\varphi^2+g_{rr}\,dr^2+g_{\theta\theta}\,d\theta^2\,,
\end{equation}
$E$, $L$, and $\Omega$ can been expressed as
\begin{eqnarray}
E&=&-\frac{g_{tt}}{\sqrt{-g_{tt}-g_{\varphi\varphi}\Omega^2}},    \label{rE}  \\
L&=&\frac{g_{\phi\phi}\Omega}{\sqrt{-g_{tt}-g_{\varphi\varphi}\Omega^2}},     \label{rL}  \\
\Omega&=&\frac{d\phi}{dt}=\sqrt{-\frac{g_{tt,r}}{g_{\varphi\varphi,r}}}.     \label{rOmega}
\end{eqnarray}
For an accretion disk around a Schwarzschild black hole pierced by cosmic string, the inner edge of the disk is at the innermost stable circular orbit(ISCO) $r_{in}=r_{ISCO}=6M/s$, which is obtained by solving the circular orbit conditions together with $dE/r=0$. The flux of the radiant energy over the disk is given as
\begin{equation}
\mathcal{F}_s(r)=\frac{3\dot{M}M^{3/2}}{8\pi s^{5/2}r^{5/2}(r s-3M)}\left[\sqrt{\frac{r s}{M}}-\sqrt{6}-\frac{\sqrt{3}}{2}\log\frac{\sqrt{r s}-\sqrt{3M}}{\sqrt{r s}+\sqrt{3M}}+\frac{\sqrt{3}}{2}\log\frac{\sqrt{2}-1}{\sqrt{2}+1}\right].
\label{eq:sch_flux}
\end{equation}
For a given parameter $s$, the flux $\mathcal{F}_s(r)$ have maximum value $\mathcal{F}_{max}$. For example, the value of $\mathcal{F}_{max}$ is  $1.146\times 10^{-4}\times\frac{3\dot{M}M^{3/2}}{8\pi}$ at $r=9.55$ as parameter $s$ take the value $s=1$.
As pointed out in Ref \cite{Luminet1979}, the observed flux $\mathcal{F}_{obs}(r)$ is different from the intrinsic flux of the source $\mathcal{F_s}(r)$ due to the redshift factor $(1+z)$. Thus, the observed flux $\mathcal{F}_{obs}(r)$ is given as \cite{Luminet1979}

\begin{equation}
\mathcal{F}_{obs}(r) = \frac{\mathcal{F}_{obs}(r)}{(1+z)^4}.
\end{equation}
\begin{figure}[h!]
    		\setlength{\tabcolsep}{ 0 pt }{\footnotesize\tt
		\begin{tabular}{ cc}
           \includegraphics[width=0.55\textwidth]{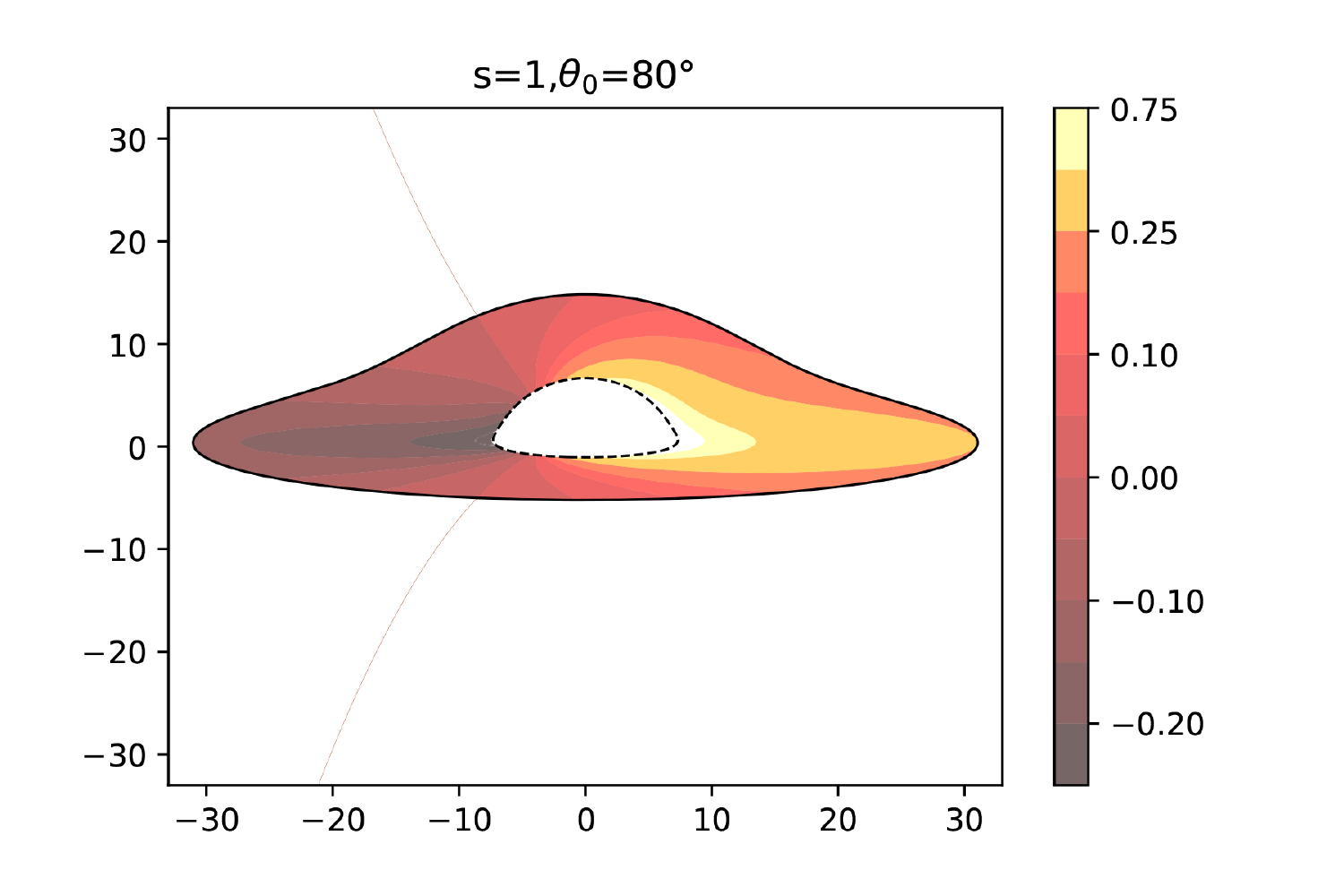}\includegraphics[width=0.55\textwidth]{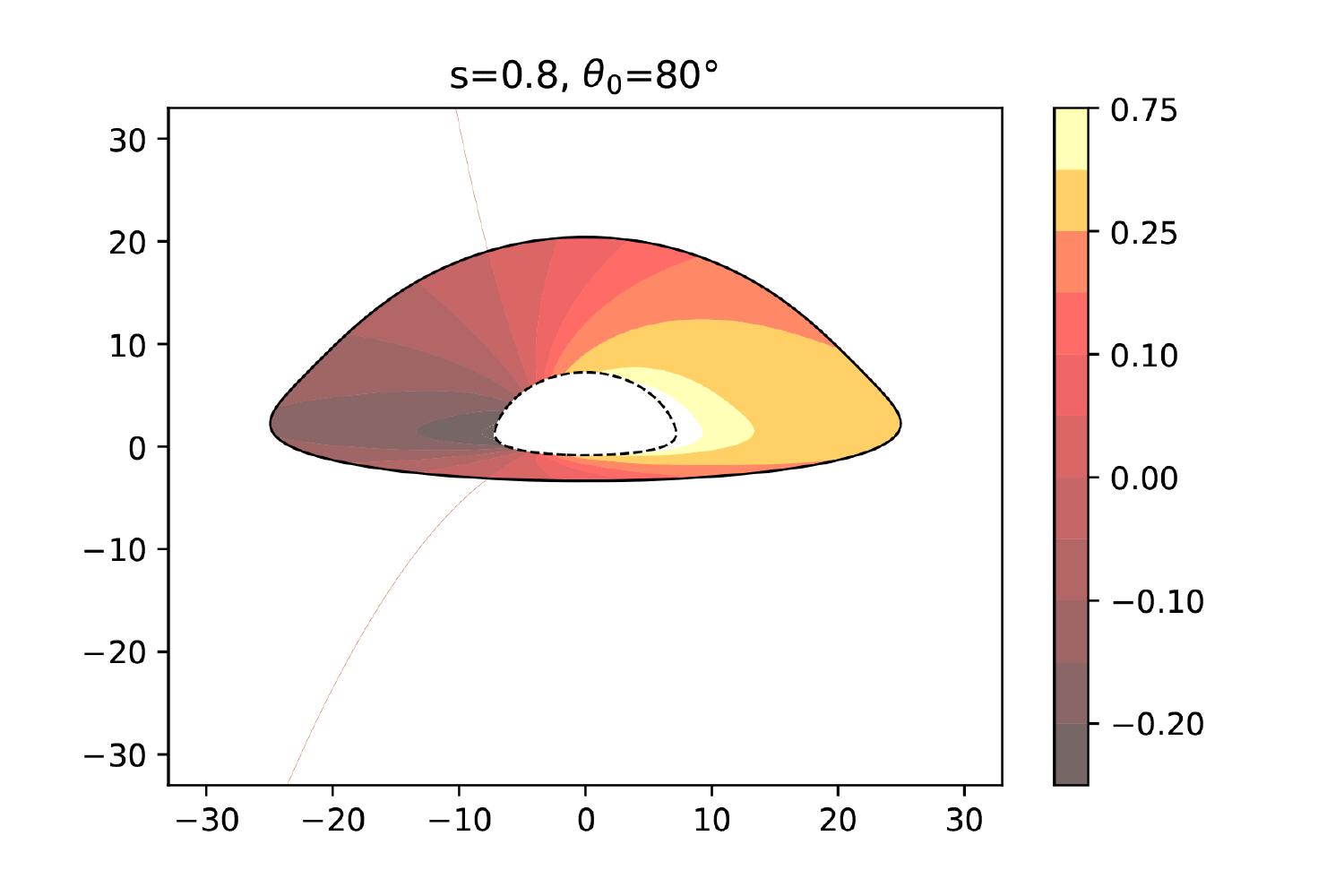}\\
           \includegraphics[width=0.55\textwidth]{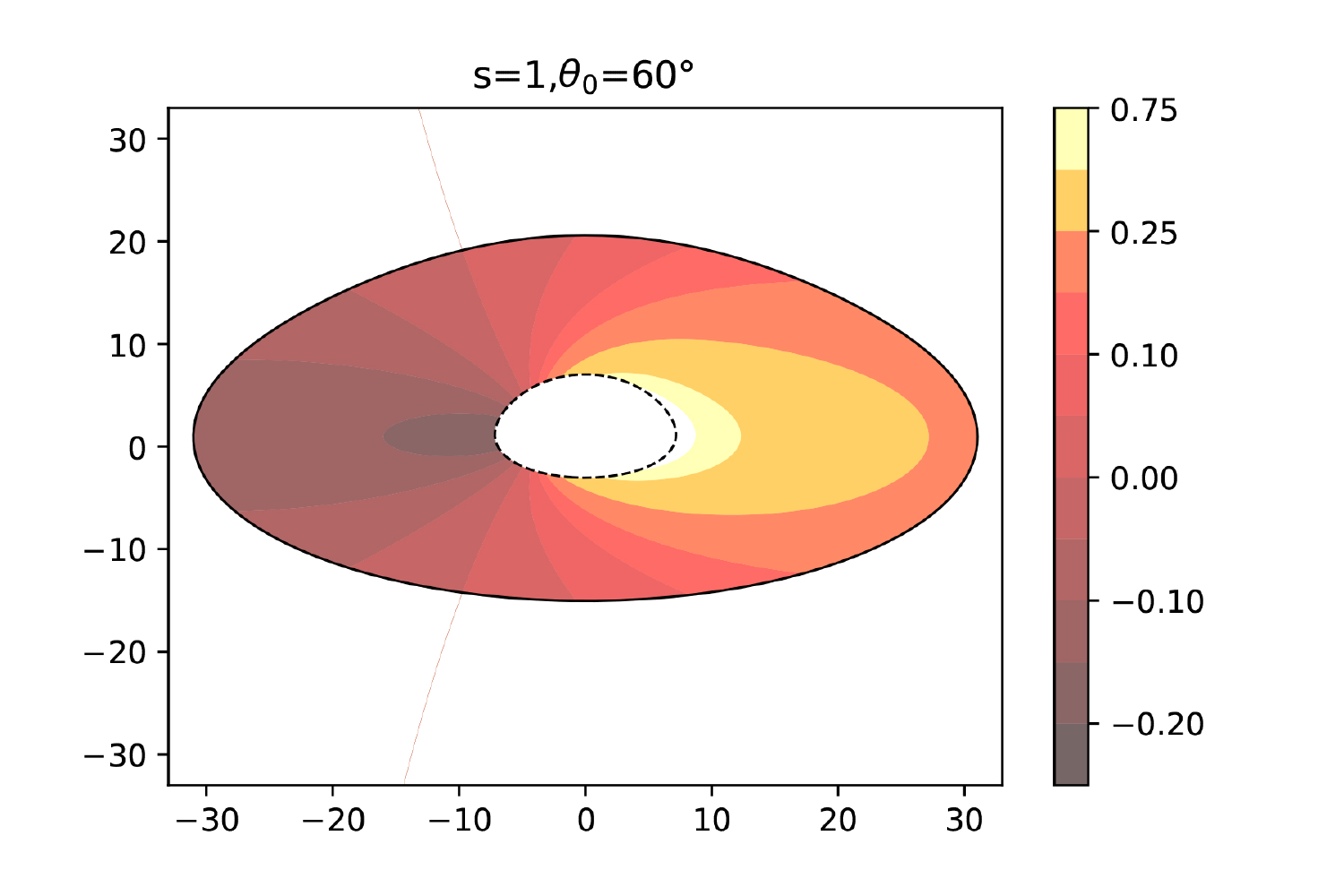}
             \includegraphics[width=0.55\textwidth]{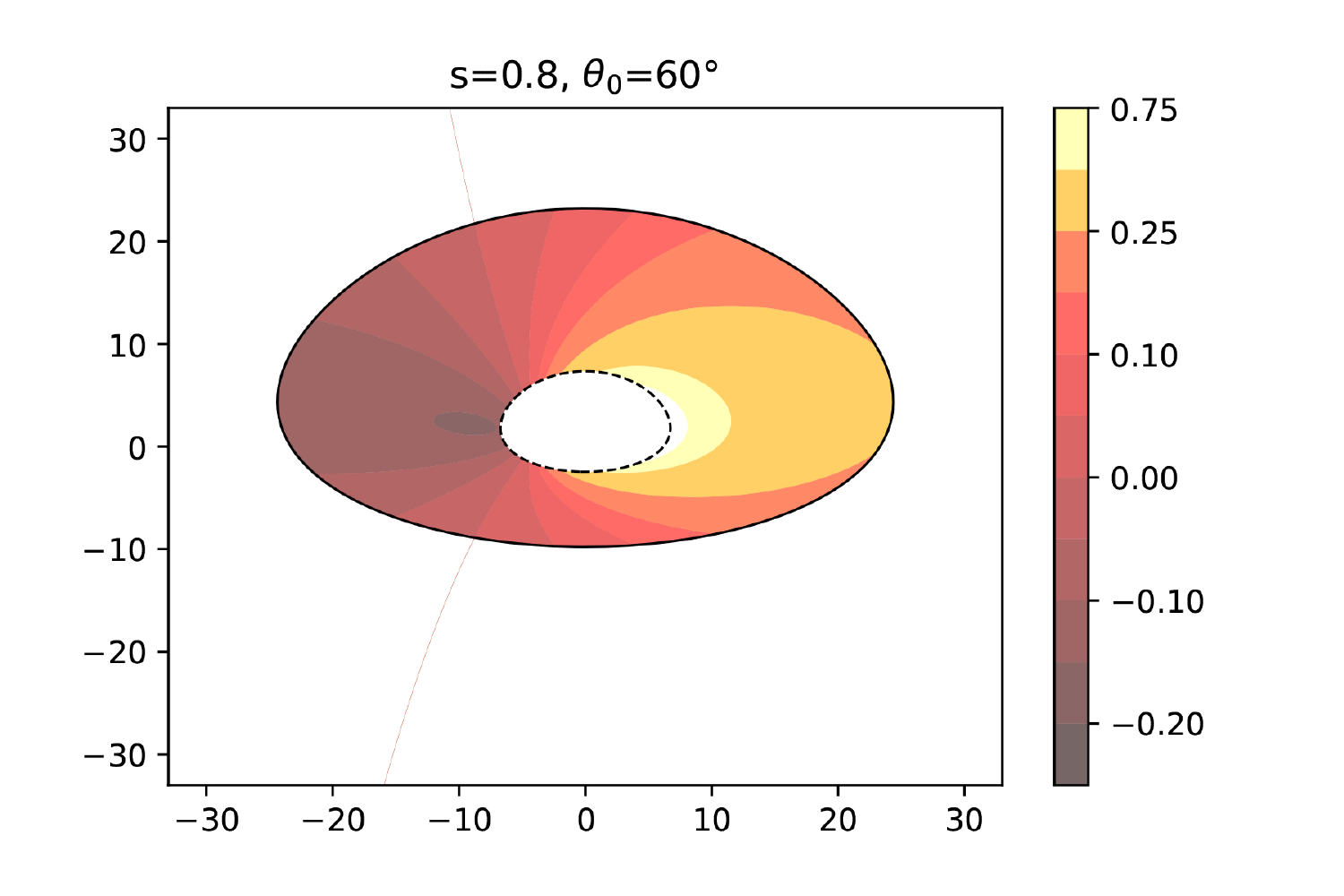}\\
             \includegraphics[width=0.55\textwidth]{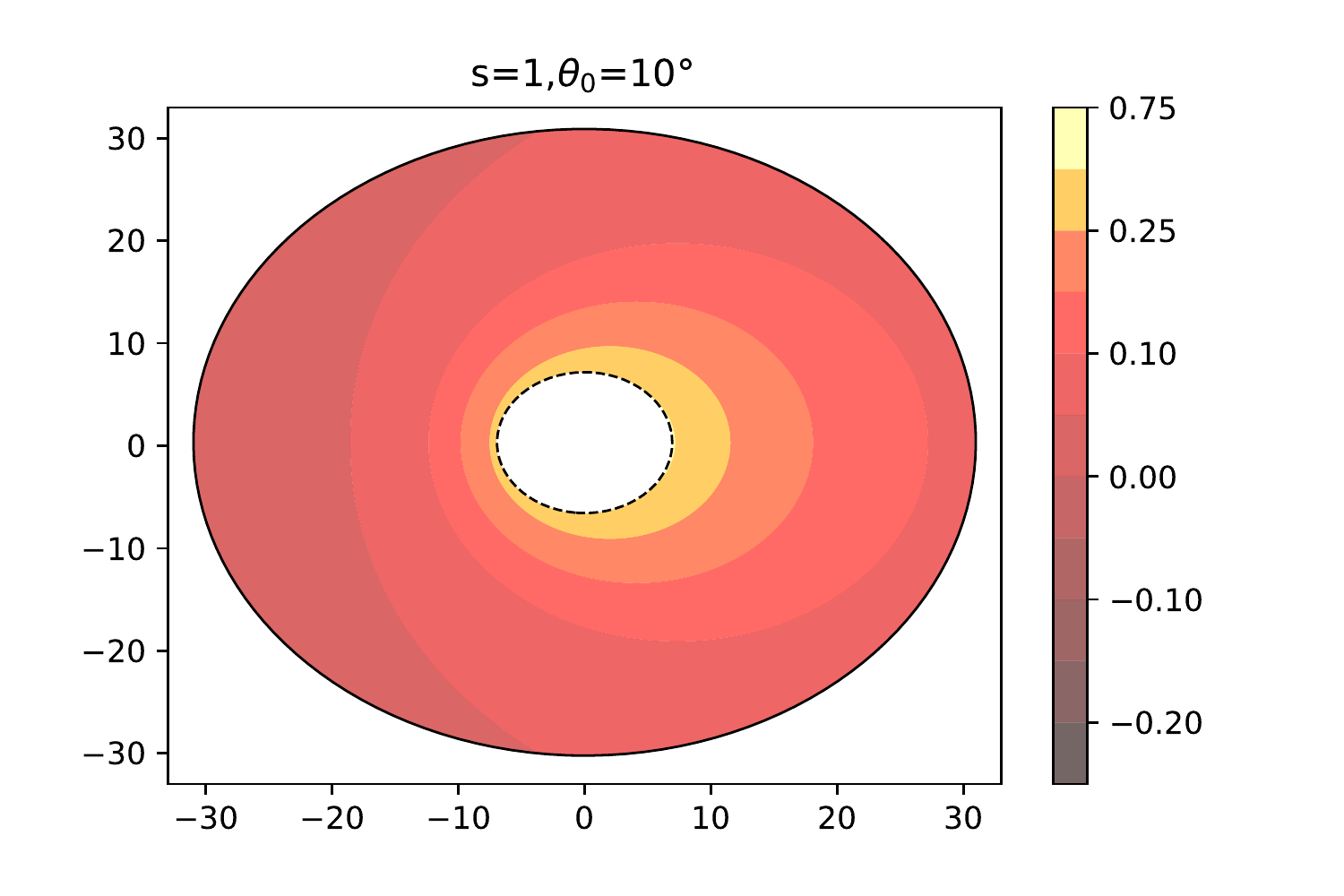}
             \includegraphics[width=0.55\textwidth]{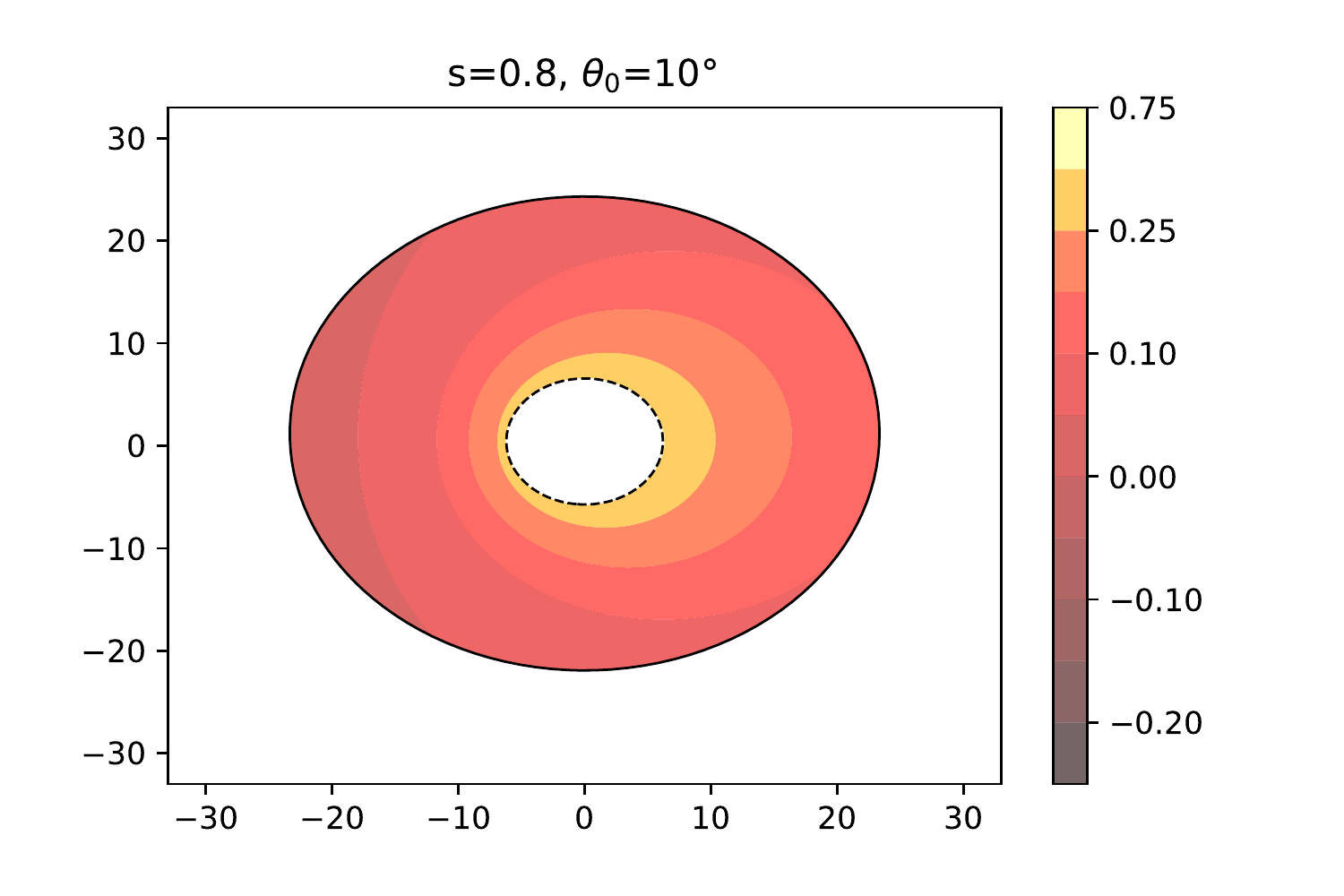}\\
		\end{tabular}}
 \caption{\label{redshift}\small Redshift distribution (curves of constant redshift $z$) of the Schwarzschild black hole pierced by cosmic string for the inclination angel $\theta_0=10^\circ, 60^\circ, 80^\circ$ ,respectively. The red solid line corresponds to
$z = 0$. The black dashed line is the inner edge of the disk at $r_{in}=r_{ISCO}$, and the outer edge of the disk is at $r=30M$. We set$M=1$.}
\end{figure}
When a photon emitted by a source particle orbiting about the black
hole, the photon has an energy which is the projection of the photon's
$4$-momentum $p$ on the 4-velocity $u$ of the emitting particle:
\begin{equation}
  E_\textrm{em}=p_\mu u^\mu=p_tu^t(1+\Omega \frac{p_\varphi}{p_t}).
\end{equation}
The components $p_t$ and $p_\Phi$ of the photon's 4-momentum are the photon's along its trajectories.
The ratio $p_t/p_\Phi$ is nothing but the impact parameter of the photon relative the $z$-axis. Using the sine theorem of spherical triangle in Fig.(\ref{coord}), we obtain
\begin{equation}\label{sb}
\frac{\sin\theta_0}{\cos\gamma}=\frac{\sin\beta}{\cos\alpha}.
\end{equation}
Thus, by Eqs. (\ref{sb}) and (\ref{eqrs}), the ratio $p_t/p_\Phi$ is given as
\begin{equation}
 \frac{p_t}{p_\varphi}=b\cos\beta=b \sin\theta_0\sin\alpha.
\end{equation}
As the photon is far from the black hole, the only nonzero component of the observer's 4-velocity is $u^t_\textrm{obs}=1$, and the observed photon's energy is $E_\textrm{obs}=p_\mu u^\mu_\textrm{obs}=p_t$. For a general static spherically symmetric metric the redshift factor is given by the expression \cite{Luminet1979}
\begin{eqnarray}\label{eq41}
  1+z&=&\frac{E_\textrm{em}}{E_\textrm{obs}}=u^t(1+\Omega b\cos\beta)\nonumber\\
  &=&\frac{(1+b\Omega\cos\beta)}{\sqrt{-g_{tt}-\Omega^2g_{\varphi\varphi}}}.
\end{eqnarray}
For the case of the Schwarzschild black hole pierced
 by a cosmic string, Eq. (\ref{eq41}) reduces to
\begin{eqnarray}\label{eq4ss}
  1+z=\frac{1+b\sqrt{\frac{M}{r^3 s^3}} \sin\alpha\sin\theta _0 }{\sqrt{1-\frac{3 M}{r s}}}.
\end{eqnarray}

\begin{figure}[h!]
    		\setlength{\tabcolsep}{ 0 pt }{\footnotesize\tt
		\begin{tabular}{ cc}
           \includegraphics[width=0.5\textwidth]{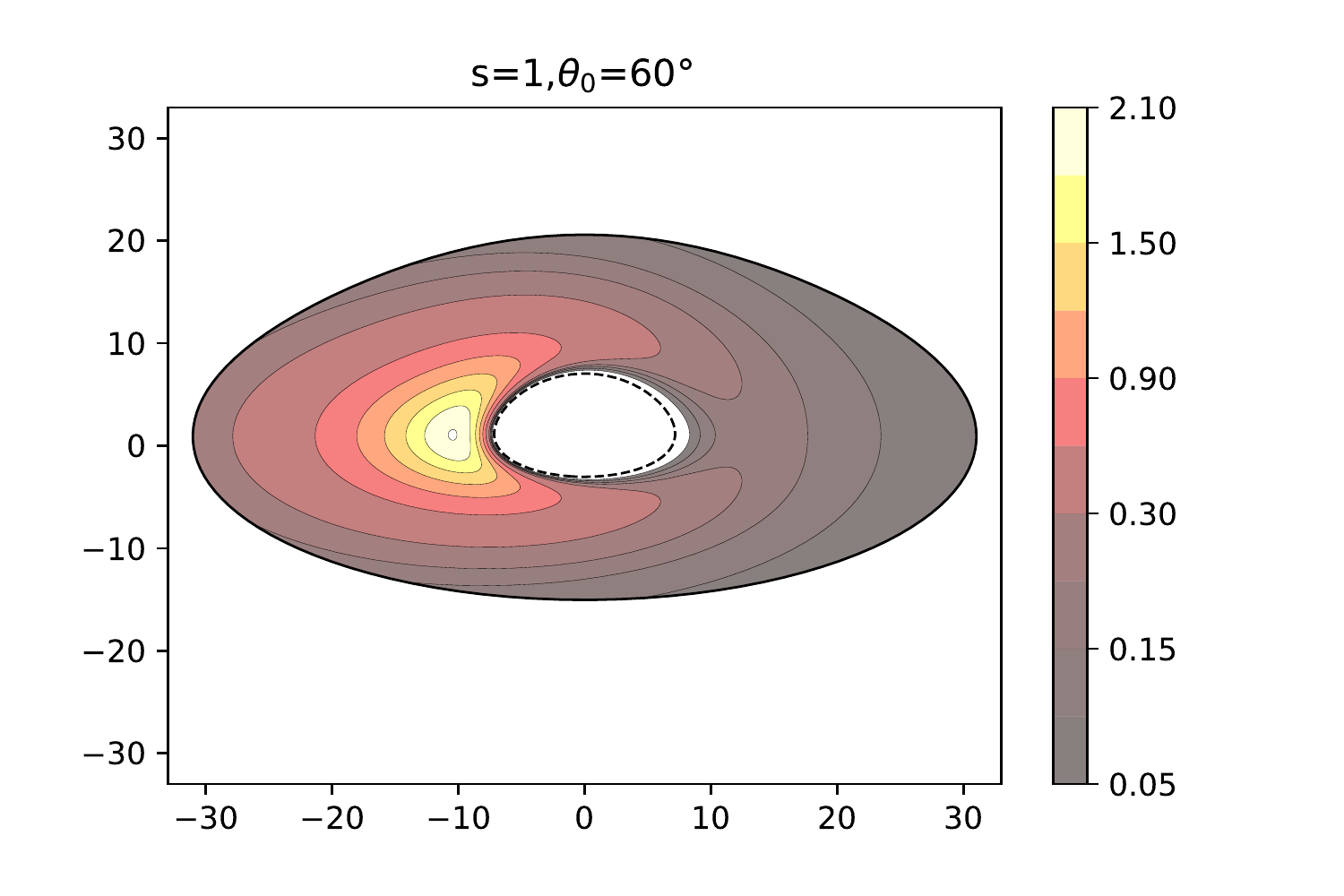}
           \includegraphics[width=0.5\textwidth]{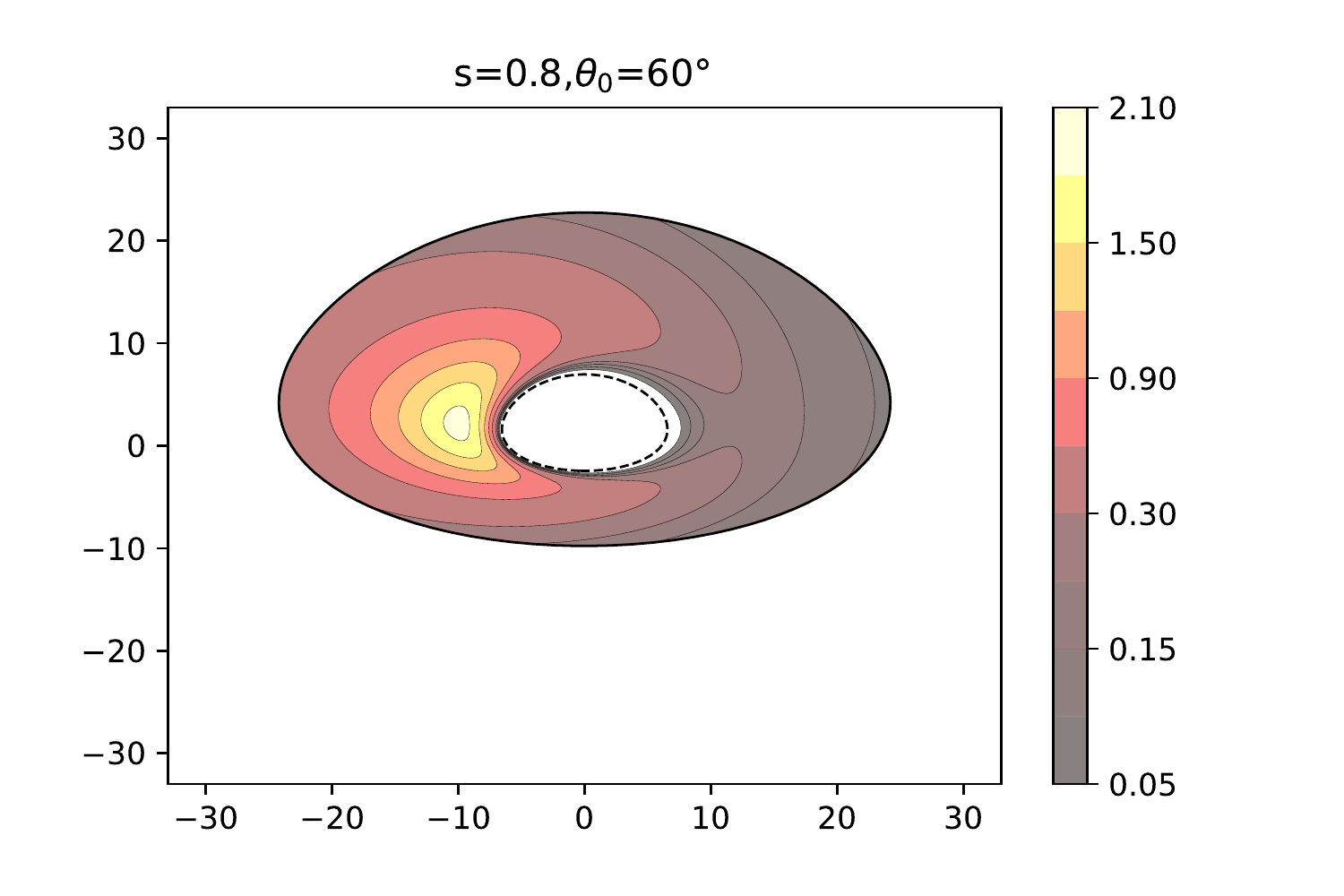}\\
           \includegraphics[width=0.5\textwidth]{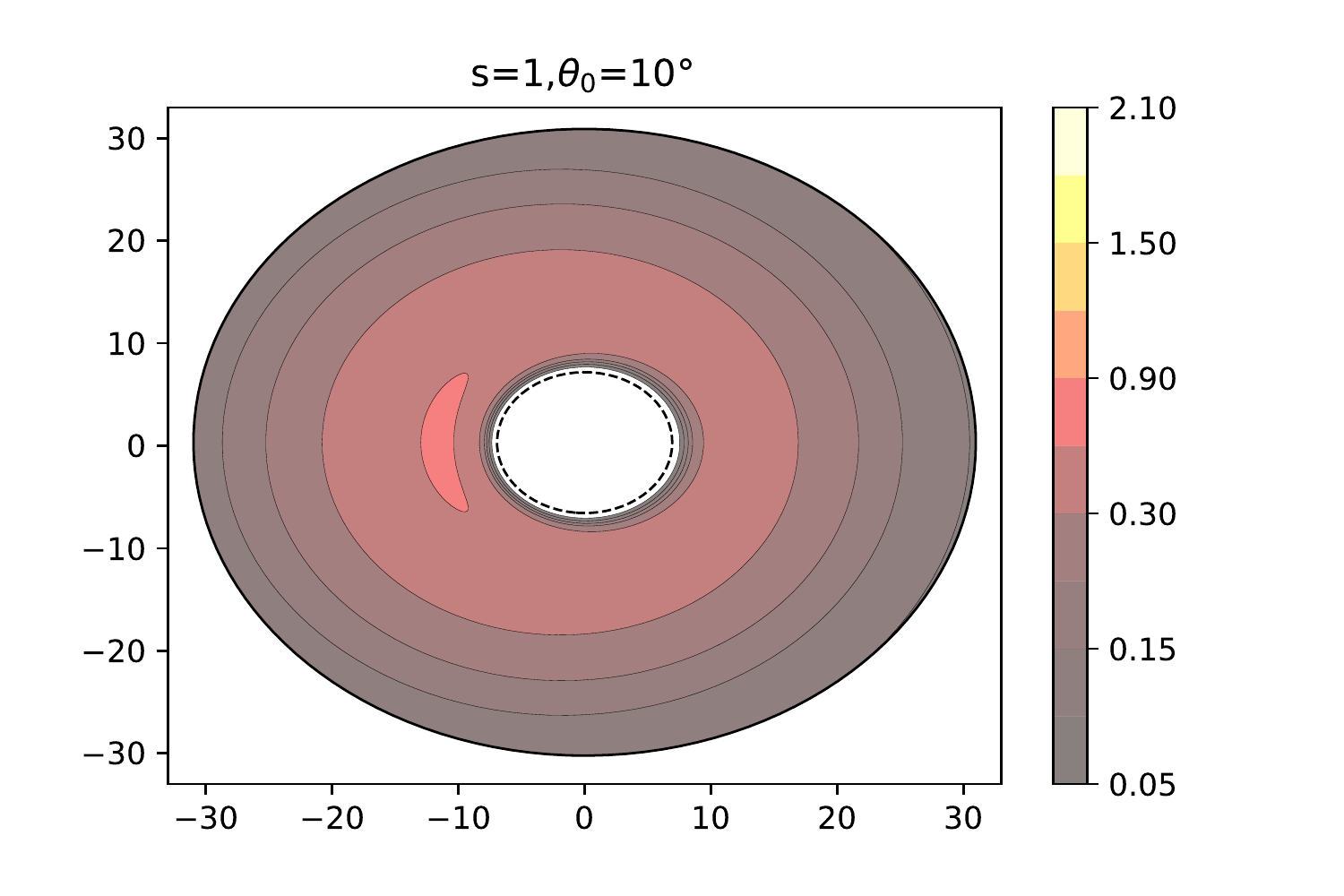}
           \includegraphics[width=0.5\textwidth]{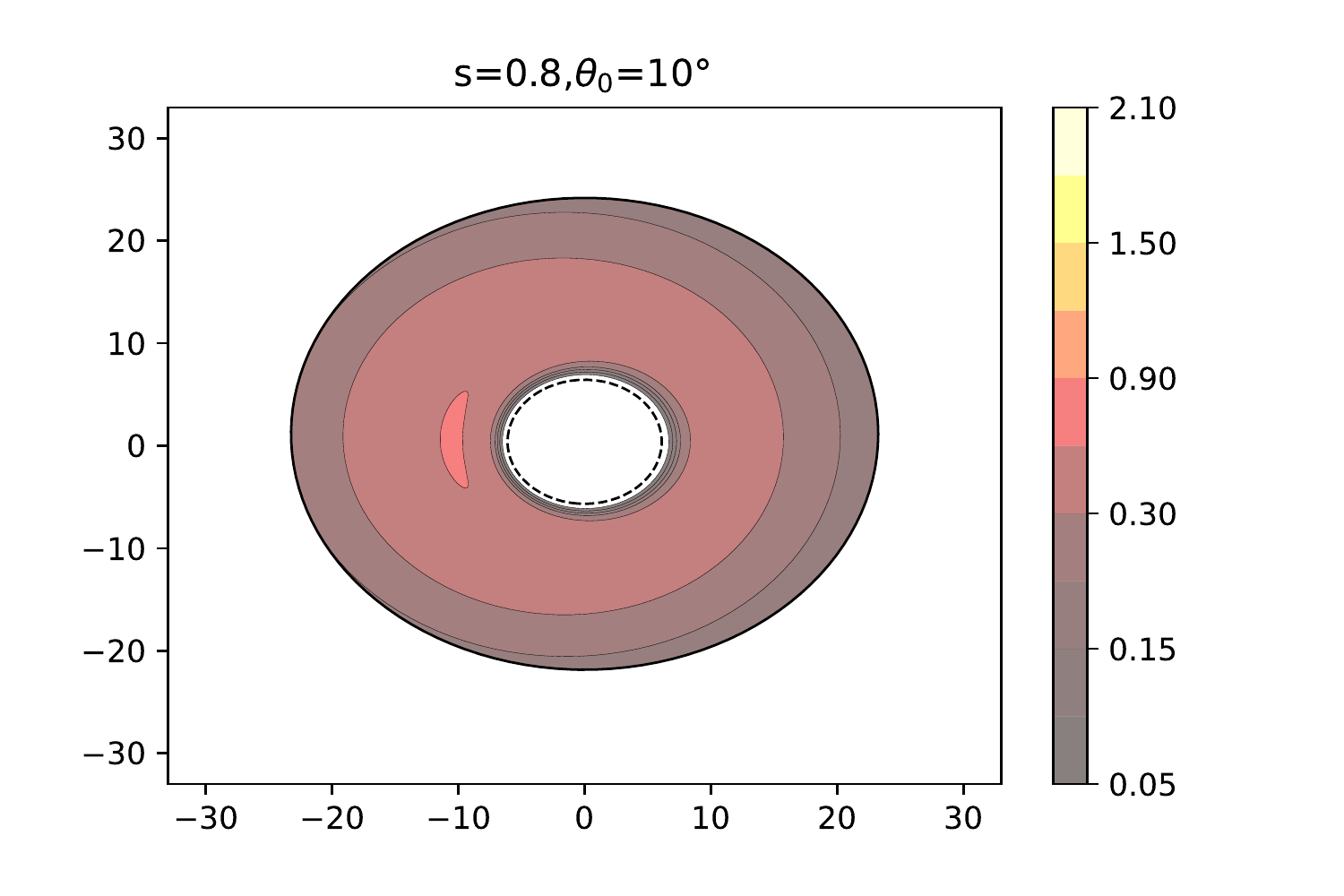}\\
         \end{tabular}}
 \caption{\label{fluxs1}\small Flux distribution in unit of $\mathcal{F}_{max}$ of direct image for the Schwarzschild black hole pierced by cosmic string with the inclination angle $\theta_0=10^\circ, 60^\circ$. The black dashed line is the inner edge of the disk at  $r_{in}=r_{isco}$, and the outer edge of the disk is at  $r=30M$. We set$M=1$. }
\end{figure}

\begin{figure}[h!]
    		\setlength{\tabcolsep}{ 0 pt }{\footnotesize\tt
		\begin{tabular}{ cc}
           \includegraphics[width=1.\textwidth]{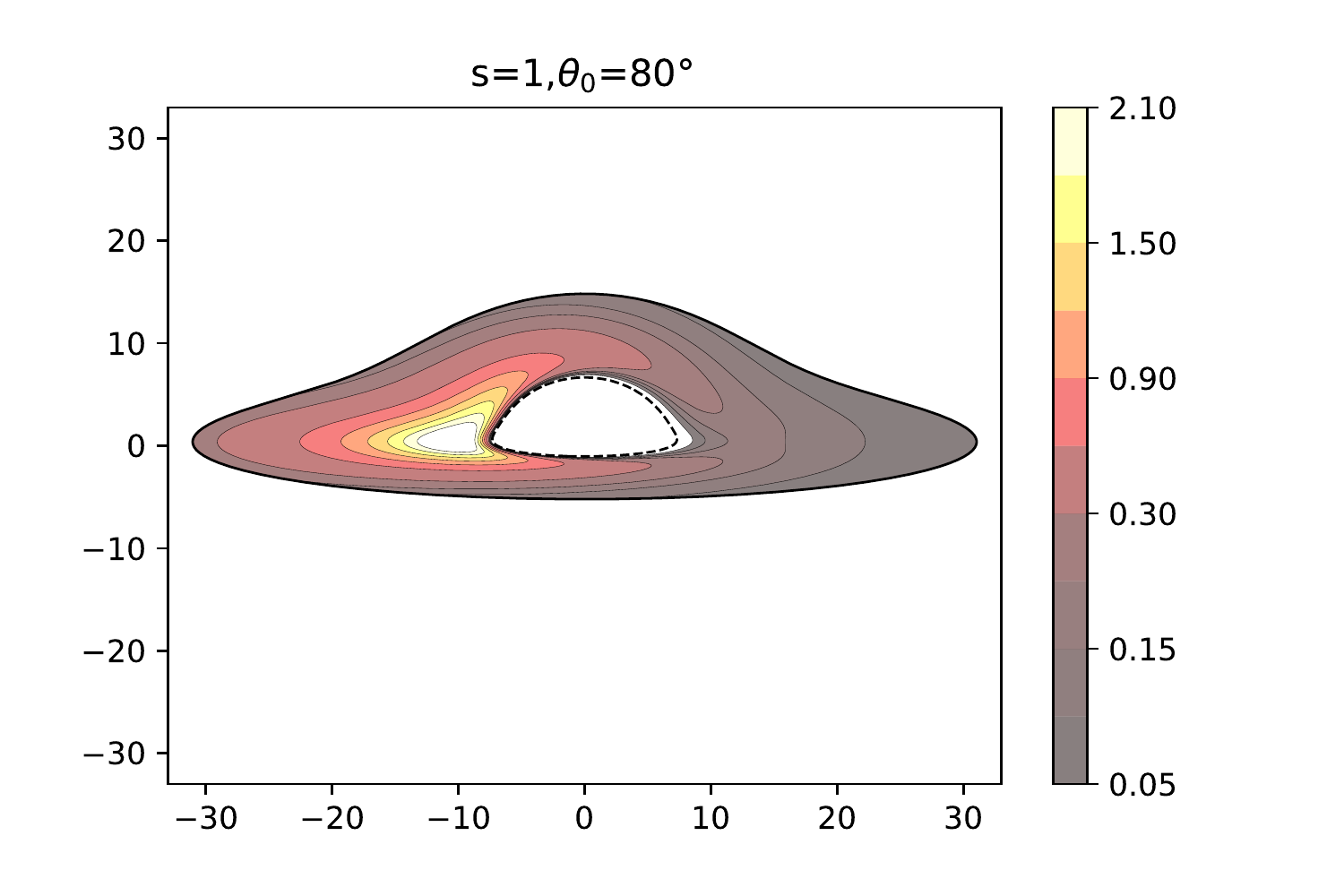}\\
           \includegraphics[width=1.\textwidth]{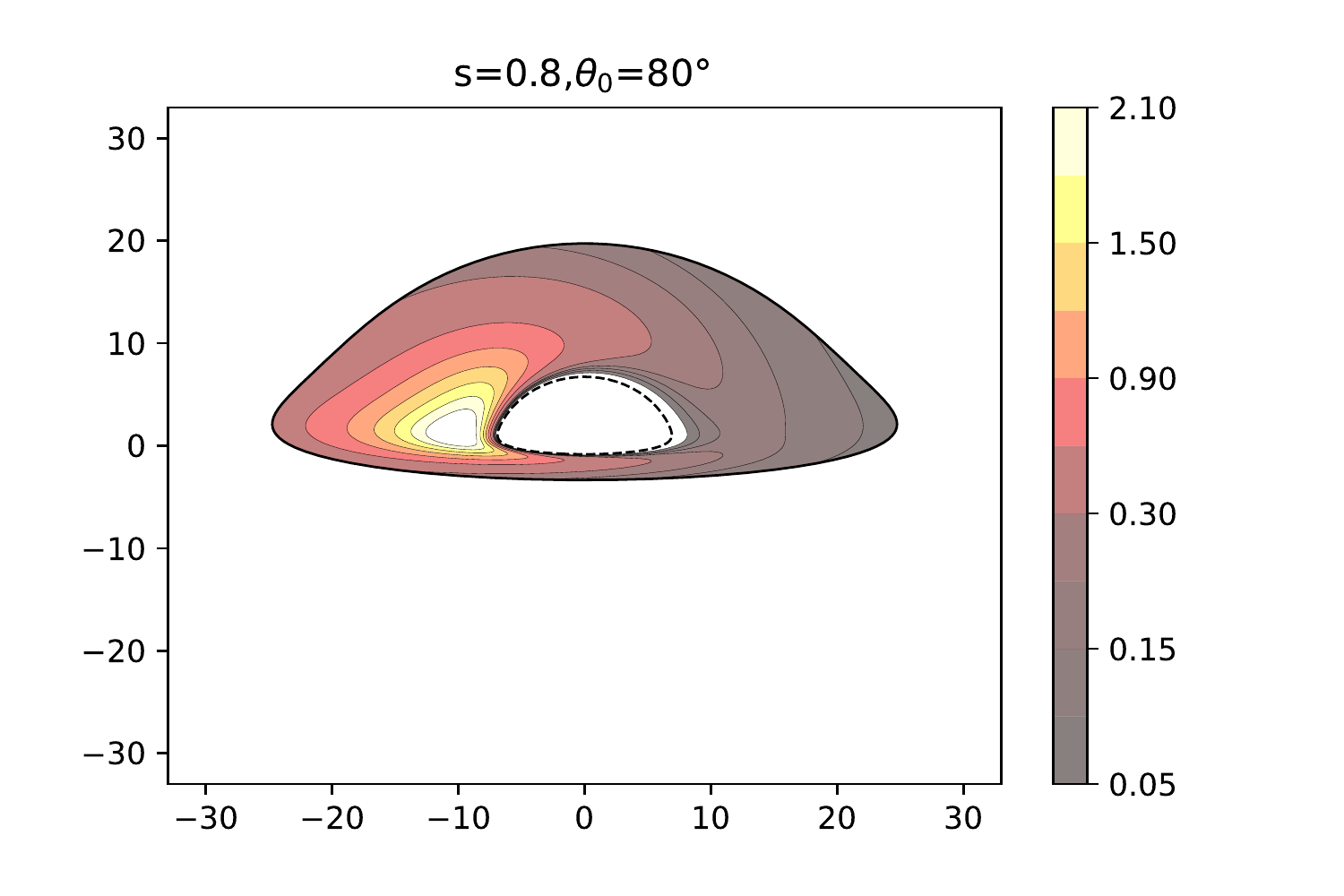}\\
         \end{tabular}}
 \caption{\label{fluxs2}\small Flux distribution in unit of $\mathcal{F}_{max}$ of direct image for the Schwarzschild black hole pierced by cosmic string with the inclination angle $\theta_0=80^\circ$. The black dashed line is the inner edge of the disk at   $r_{in}=r_{isco}$, and the outer edge of the disk is at  $r=30M$. We set$M=1$. }
\end{figure}

\begin{figure}[h!]
    		\setlength{\tabcolsep}{ 0 pt }{\footnotesize\tt
		\begin{tabular}{ cc}
           \includegraphics[width=0.5\textwidth]{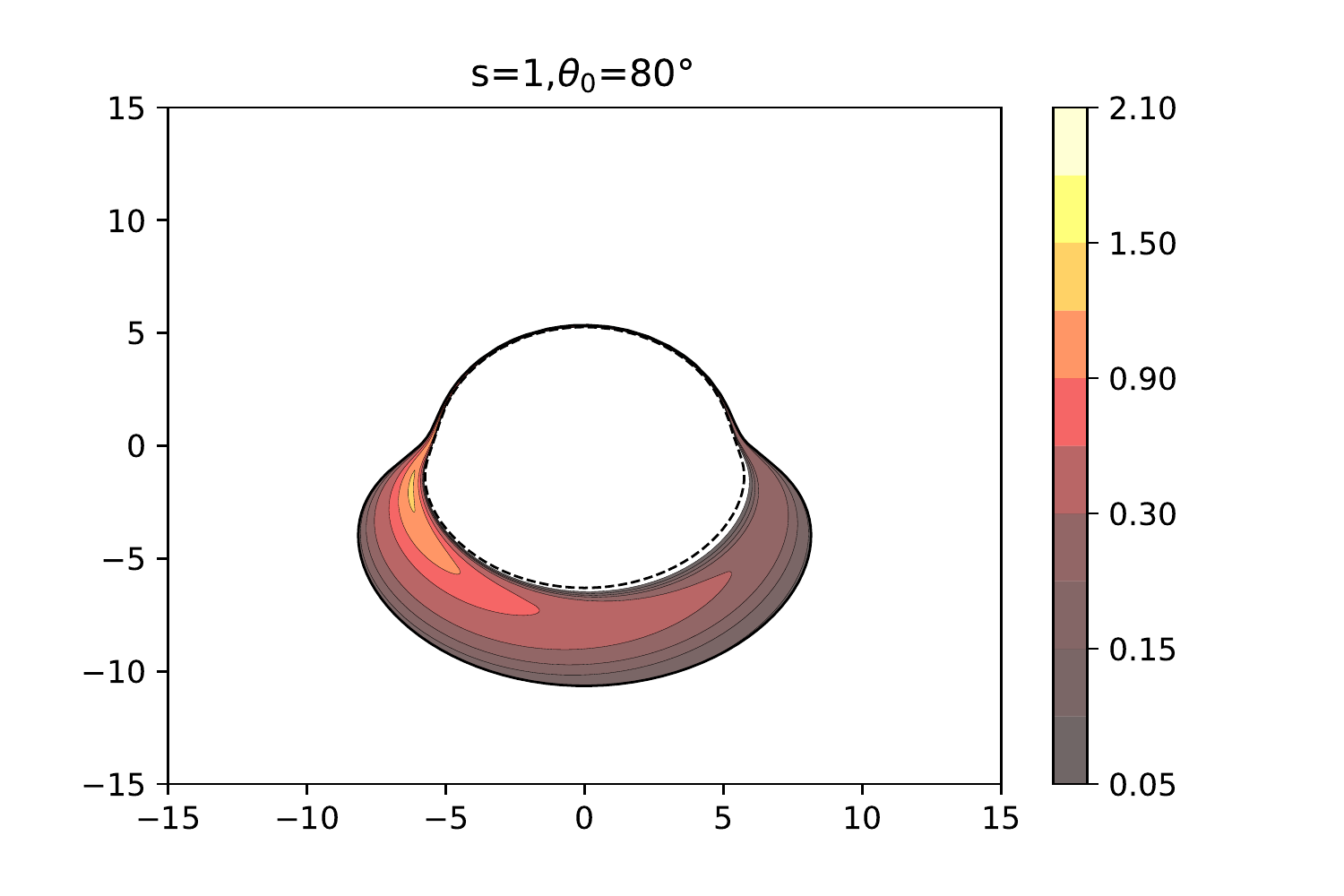}\includegraphics[width=0.5\textwidth]{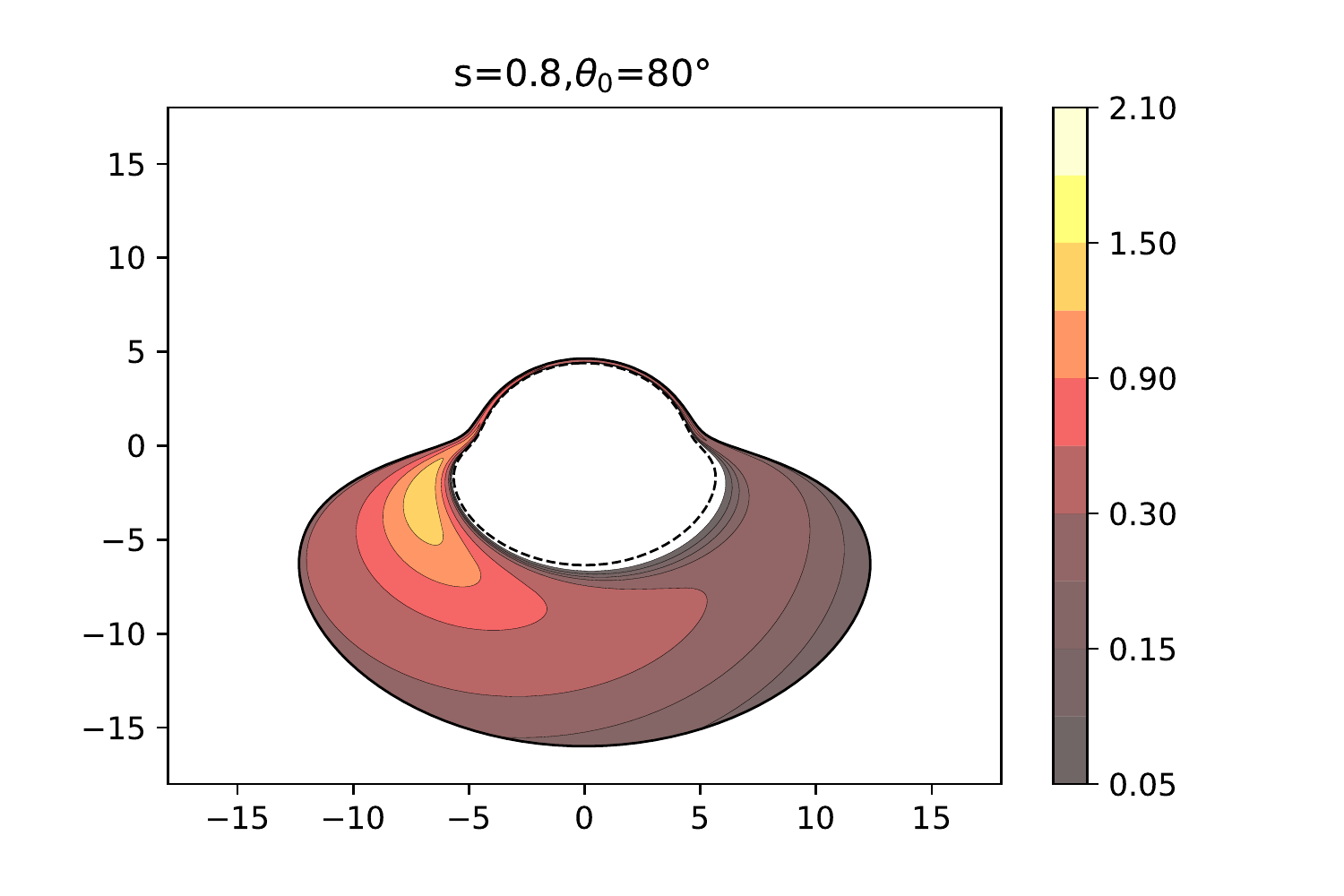}\\
            \includegraphics[width=0.5\textwidth]{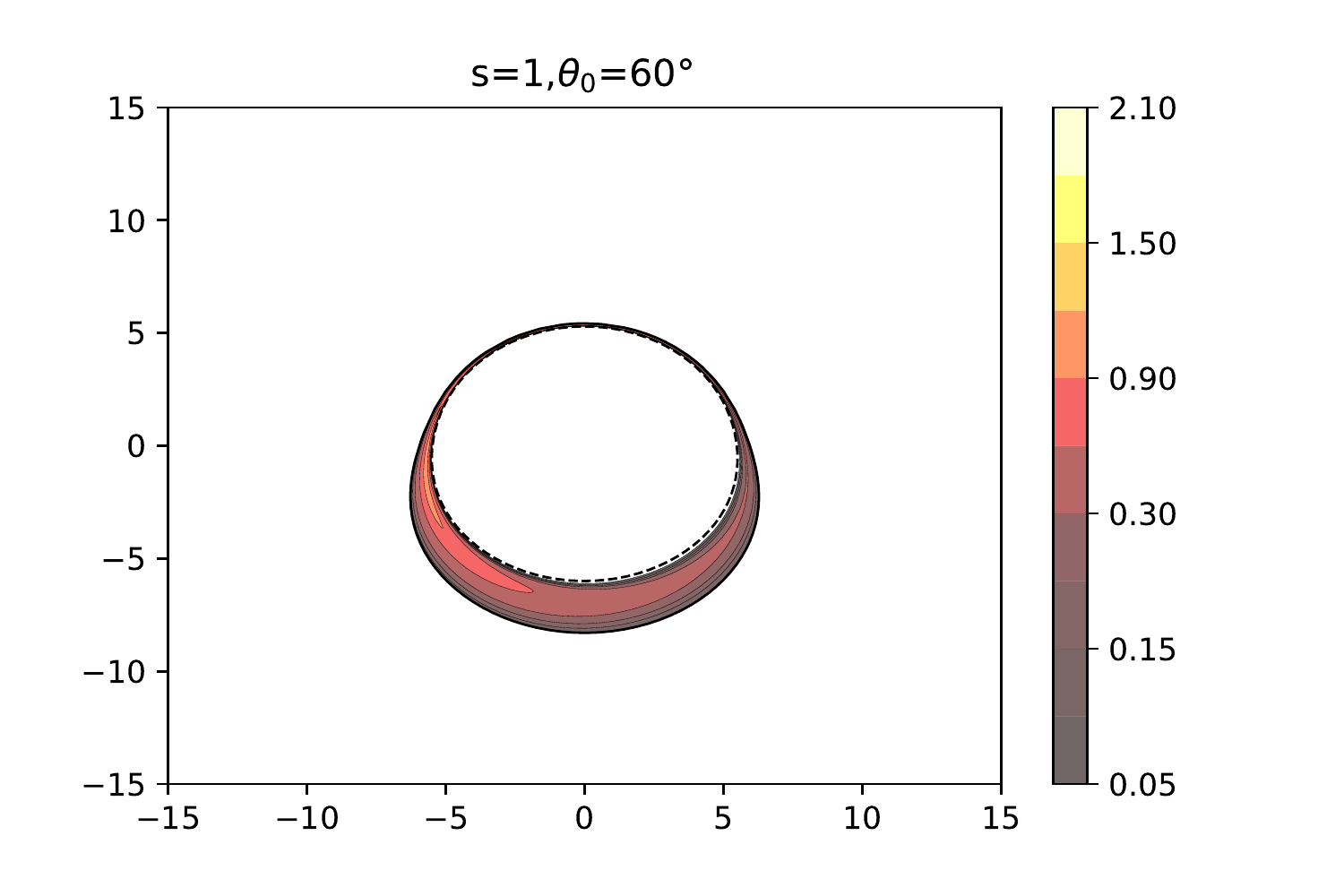}
           \includegraphics[width=0.5\textwidth]{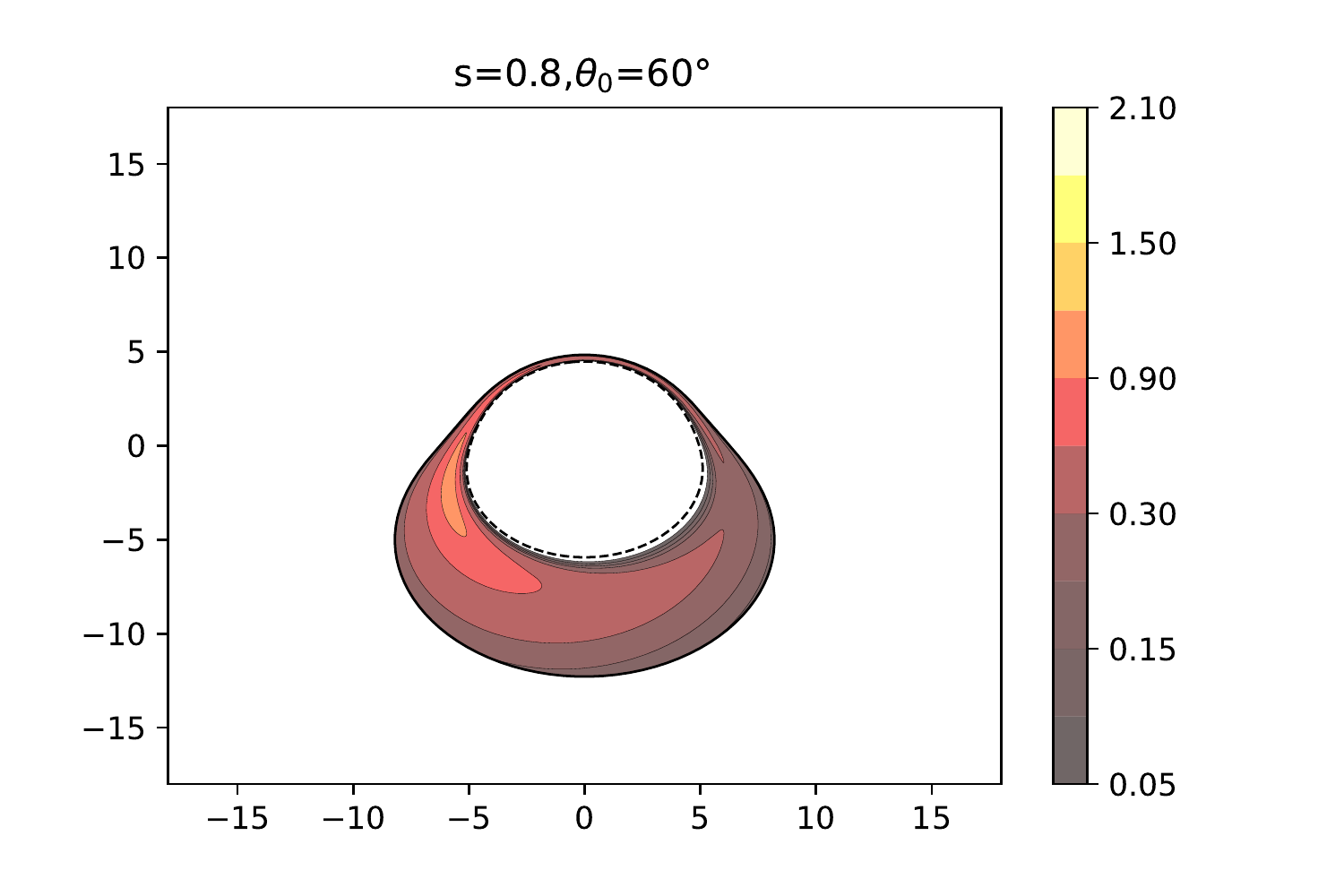}\\
             \includegraphics[width=0.5\textwidth]{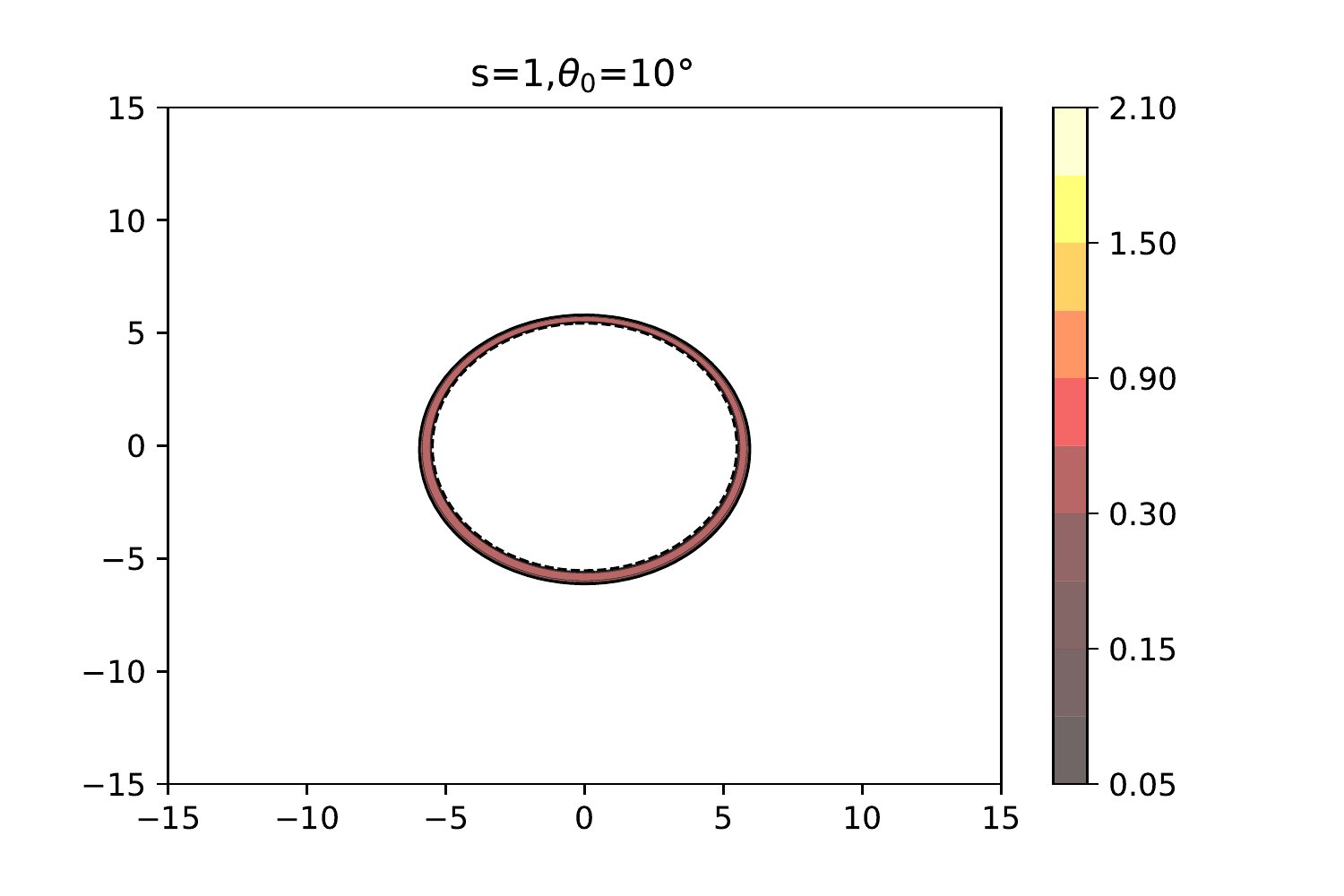}
           \includegraphics[width=0.5\textwidth]{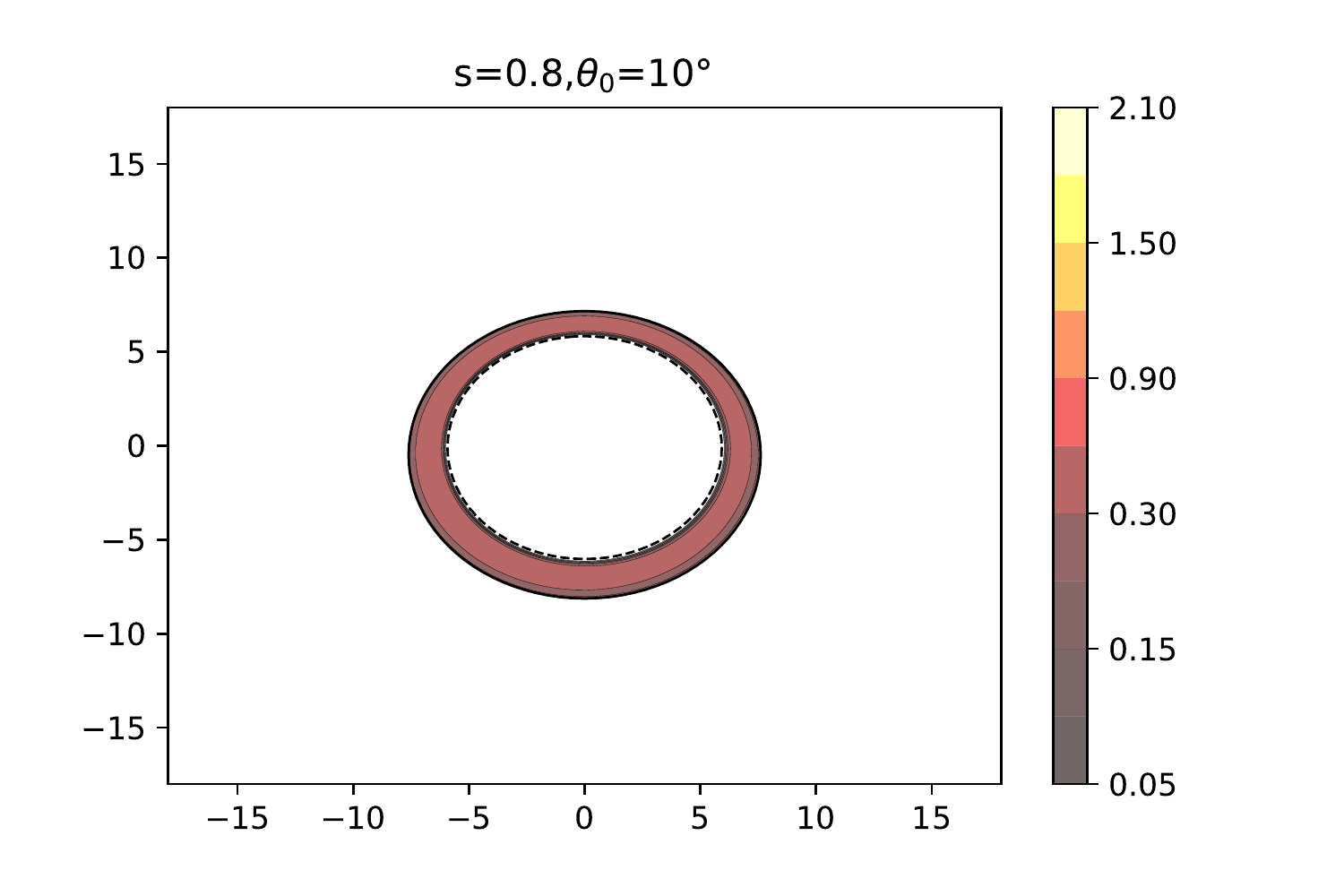}\\[1mm]
		\end{tabular}}
 \caption{\label{fluxs3}\small Flux distribution in unit of $\mathcal{F}_{max}$ of secondary image for the Schwarzschild black hole pierced by cosmic string with the inclination angle $\theta_0=10^\circ, 60^\circ,80^\circ$. The  black dashed line is the inner edge of the disk at $r_{in}=r_{isco}$, and the outer edge of the disk  is at $r=30M$. We set$M=1$}.
\end{figure}

Figure. \ref{redshift} shows redshift distribution (curves of constant redshift $z$) in the direct image for the inclination angel $\theta_0=10^\circ,60^\circ, 80^\circ$ and the cosmic string parameters $s=1, 0.8$, respectively. It is ease see that, for the high inclination angles $\theta_0=60^\circ, 80^\circ$,
blueshift($z<0$ due to rotation of the disk) in the left half-part of the plate exceeds the gravitational redshift($z>0)$ part due to the the existence of the black hole. On the other hand, for the low inclination angle $\theta_0=10^\circ$, blueshift distribution does not exist. We also find that the region of redshift distribution with high redshift value $z$ increases with the decrease of cosmic string parameter $s$.

Figures.\ref{fluxs1}, \ref{fluxs2} and \ref{fluxs3} present flux distribution(curves of constant  $\mathcal{F}_{obs}$) in unit of $\mathcal{F}_{max}$ of direct and secondary image with the inclination angle $\theta_0=10^\circ, 60^\circ,80^\circ$ and the cosmic string parameter $s=1,0.8$, respectively. The flux is of course maximum in the regions where the spectral shift is blueshift. The flux distributions become strongly asymmetric with the increase of the observer's inclination angle $\theta_0$. The region of the flux distribution with large flux values $\mathcal{F}_{obs}/\mathcal{F}_{max}$ in the direct(secondary) image become smaller(larger) with the decrease of cosmic string parameter $s$.

\section{Conclusion}\label{sec3}

In this paper, we have studied the optical appearance of a thin accretion disk surrounding the Schwarzschild black hole pierced by cosmic string and its observable radiation in the framework of the Novikov-Thorne model. Using the semi-analytical method of Luminet\cite{Luminet1979}, we expressed the solutions to geodesic equations in terms of elliptic integrals. By inverting their expressions into Jacobi elliptic functions, iso-radial curves corresponding to photons emitted at a constant radius from the black hole as seen by a distant observer in different inclination angles have been plotted in Fig \ref{images}. We treat those iso-radial curves as the direct and secondary images of the thin disc. Those images are symmetric with respect to
the axis defined by the normal to orbital plane. This is a common feature of static spherically symmetric black holes, however, it does not hold for steady-state rotating black holes\cite{Cunningham:1975}. At the same time, Fig \ref{images} also show that the cosmic string parameter $s$ can modify the shape and size of both the direct and secondary images. Finally, we calculated the value of redshift $z$ and the observed flux at each point in the observer's photographic plate at polar coordinates $(b, \alpha )$.
The redshift distributions of the direct images were plotted in Fig. \ref{redshift} and the flux distributions of both the direct and secondary images were depicted in Figs.\ref{fluxs1}, \ref{fluxs2} and \ref{fluxs3}. It is shown that the
cosmic string parameter $s$ indeed has an effect on the optical appearance of the holes. Our result may provide some insight imprints of cosmic string on the images of the black holes for the astronomical observation with EHT and GRAVITY
in the future.

\section*{Acknowledgments}
 This work was supported by the
Scientific Research Fund of the Hunan Provincial Education Department under No. 19A260, the National
Natural Science Foundation (NNSFC) of China (grant No. 11447168).

\end{document}